\documentclass[aps,prd,twocolumn,showpacs,nofootinbib,notitlepage,eqsecnum,
superscriptaddress]{revtex4-1}

\usepackage[centertags]{amsmath}
\usepackage{amssymb}
\usepackage{latexsym}
\usepackage{enumerate}
\usepackage{enumitem}
\usepackage{graphicx}
\usepackage{mathrsfs}
\usepackage{stmaryrd}
\usepackage{import}
\usepackage{tensor}
\usepackage{color}
\usepackage{bm}
\usepackage{multirow}
\usepackage{breakurl}
\usepackage{float}
\usepackage{placeins}
\usepackage{autobreak}
\usepackage{array}
\usepackage{mathtools}
\usepackage[table,dvipsnames]{xcolor}
\usepackage{pbox}
\usepackage{hyperref}

\newcommand{\bi}{\begin{itemize}}
\newcommand{\ei}{\end{itemize}}
\newcommand{\be}{\begin{equation}}
\newcommand{\ee}{\end{equation}}

\newcommand{\q}{\quad}
\newcommand{\qq}{\qquad}

\newcommand{\vp}{\varphi}

\allowdisplaybreaks

\begin{document}

\title{Eccentric-orbit EMRI radiation: Analytic forms of leading-logarithm and subleading-logarithm 
flux terms at high PN orders}

\author{Christopher Munna}
\author{Charles R. Evans}
\affiliation{Department of Physics and Astronomy, University of North 
Carolina, Chapel Hill, North Carolina 27599, USA}

\begin{abstract}
We present new results on the analytic eccentricity dependence of several 
sequences of gravitational wave flux terms at high post-Newtonian (PN) order 
for extreme-mass-ratio inspirals.  These sequences are the leading logarithms, 
which appear at PN orders $x^{3k} \log^k(x)$ and $x^{3k+3/2} \log^k(x)$ for 
integers $k\ge 0$ ($x$ a PN compactness parameter), and the subleading 
logarithms, which appear at orders $x^{3k} \log^{k-1}(x)$ and 
$x^{3k+3/2} \log^{k-1}(x)$ ($k\ge 1$), in both the energy and angular momentum
radiated to infinity. For the energy flux leading logarithms, we 
show that to arbitrarily high PN order their eccentricity dependence is 
determined by particular sums over the function $g(n,e_t)$, derived from the 
Newtonian mass quadrupole moment, that normally gives the spectral content of 
the Peters-Mathews flux as a function of radial harmonic $n$.  An analogous 
power spectrum $\tilde{g}(n,e_t)$ determines the leading logarithms of the
angular momentum flux.  For subleading logs, the quadrupole power spectra are 
again shown to play a role, providing a distinguishable part of the 
eccentricity dependence of these flux terms to high PN order.  With the 
quadrupole contribution understood, the remaining analytic eccentricity 
dependence of the subleading logs can in principle be determined more easily 
using black hole perturbation theory.  We show this procedure in action, 
deriving the complete analytic structure of the $x^6 \log(x)$ subleading-log 
term and an analytic expansion of the $x^{9/2}$ subleading log to high order 
in a power series in eccentricity.  We discuss how these methods might be 
extended to other sequences of terms in the PN expansion involving logarithms. 
\end{abstract}


\maketitle

\section{Introduction}
\label{sec:intro}

With gravitational wave observations of merging compact binaries by LIGO and Virgo \cite{AbboETC16a,BaraETC18} 
now routine, researchers look forward to the LISA mission \cite{eLISA,LISA} and eventual detection 
of new classes of events, such as extreme-mass-ratio inspirals (EMRIs) that involve a stellar mass black hole 
($\mu \sim 10 M_\odot$) spiralling toward a supermassive black hole ($M \sim 10^6 M_\odot$).  For EMRIs the 
small mass ratio $\varepsilon \coloneqq \mu/M \ll 1$ serves as a perturbation parameter allowing the 
Einstein equations to be solved in an expansion in powers of $\varepsilon$.  In this black hole perturbation theory 
(BHPT) approach, the back-reaction on the small body's motion requires calculation of the regularized gravitational 
self-force (GSF) \cite{PoisPounVega11}.  Recent progress in this area has included first-order long-term inspiral 
calculations \cite{OsbuWarbEvan16,WarbOsbuEvan17} of EMRIs with a nonspinning primary and calculation of the 
first-order GSF for generic orbits about a spinning (Kerr) primary \cite{vdMe17}.

Post-Newtonian (PN) theory, alternatively, is best suited for wide orbits and slow orbital motions $v/c\ll1$, or 
equivalently for small (dimensionless) orbital frequencies, where $x\coloneqq ((m_1 + m_2)\Omega_\varphi)^{2/3}\ll 1$) 
is a compactness parameter \cite{Blan14}.  Peters and Mathews \cite{PeteMath63,Pete64} were first to calculate 
eccentric binary evolution subject to gravitational radiation at lowest PN order (i.e., quadrupole radiation).  
Modeling general orbits is important, as EMRIs are expected to have moderate to high eccentricities 
\cite{BaraCutl04,HopmAlex05,AmarETC07}.  For nonspinning compact binaries, the gravitational wave phase has now 
been calculated to 3PN order \cite{ArunETC08a,ArunETC08b,ArunETC09a} for eccentric orbits and 3.5PN order 
\cite{MoorETC16} for quasi-circular orbits.  The equations of motion have been extended to 4PN order 
(see \cite{BaraETC18} for a review).  

These two approaches to the two-body problem overlap for EMRIs that are early in an inspiral, and considerable 
research has proceeded in recent years cross-checking results from the two techniques (thus far almost exclusively 
at first order in the mass ratio) and uncovering the PN expansion of BHPT/GSF quantities.  Initially, analytic terms 
in the PN expansions were determined through inspection of accurate BHPT/GSF numerical results.  The earliest example 
of this procedure was the recognition that $4\pi$ matched the numerical coefficient seen in BHPT calculations 
\cite{CutlETC93} of the 1.5PN tail in the energy flux for circular orbits, with the result being separately 
confirmed theoretically \cite{Wise93,Pois93}.  Later, starting with Detweiler \cite{Detw08}, efforts were made 
\cite{SagoBaraDetw08,BaraSago09,BlanETC09,BlanETC10,ShahFrieWhit14,DolaETC14b,JohnMcDaShahWhit15,
BiniDamoGera15,AkcaETC15,HoppKavaOtte16,BiniDamoGera16a,BiniDamoGera16b, BiniDamoGera16c} 
to identify analytic terms in the PN expansion for gauge-invariant quantities in the conservative GSF sector, such as
the redshift invariant $u^t$.  Similar progress has been made in finding analytic coefficients at high PN order for the 
fluxes from circular orbits \cite{Fuji12a,Fuji12b,Shah14} and eccentric orbits \cite{SagoFuji15,ForsEvanHopp16}. 

Part of these efforts involved development of extreme high-accuracy (e.g., hundreds of decimals of accuracy) 
BHPT and GSF \textsc{Mathematica} calculations \cite{ShahFrieWhit14,Shah14,JohnMcDaShahWhit15,ForsEvanHopp16} centered 
around use of the MST (Mano-Suzuki-Tagasuki) analytic function expansion formalism 
\cite{ManoSuzuTaka96a,ManoSuzuTaka96b,SasaTago03}.  Numerical results from different orbital radii (as well as 
eccentricity \cite{ForsEvanHopp16}) are fitted to the form of an expected PN expansion to determine coefficients 
numerically.  Then the high accuracy of the floating point numbers allows an integer relation algorithm (PSLQ) 
\cite{FergBailArno99} to ferret out the underlying rational and transcendental numbers that make up these 
coefficients.  It was subsequently realized that \textsc{Mathematica} codes might directly calculate 
\cite{BiniDamo13,BiniDamo14a,BiniDamo14b,BiniDamo14c,KavaOtteWard15,HoppKavaOtte16} the 
PN expansion of the MST solutions and store and output massively long 
expressions for BHPT/GSF quantities for arbitrary orbital parameters, rather than evaluate numerical values for 
specific orbits.  The results in this paper were in some cases checked and in other cases derived by using both a 
high-precision numerical MST code and a new all-analytic code.

The present paper and one being written contemporaneously \cite{MunnETC19} concern the analytic form of the 
PN expansion for gravitational wave fluxes to infinity from eccentric nonspinning EMRIs.  Drawing upon an earlier 
effort \cite{ForsEvanHopp16}, the companion paper \cite{MunnETC19} significantly extends the analytic understanding 
of energy flux between 3.5PN and 9PN order, as a simultaneous expansion in PN order and powers of the eccentricity 
$e$, and presents the equivalent explication of angular momentum flux.  This paper focuses on two subsets of PN terms 
called leading logarithms \cite{GoldRoss10} and subleading logarithms and uses a mix of PN analysis and examination 
of BHPT results to provide a theoretical understanding of the eccentricity functional dependence of these 
logarithmic terms.  

Leading logarithms are a sequence that appear at PN orders $x^{3k} \log^k(x)$ and $x^{3k+3/2} \log^k(x)$ for 
integers $k\ge 0$.  (Here and henceforth in this paper PN order in the fluxes refers to order relative to lowest 
order quadrupole radiation.)  Leading logs are defined as those terms in which a new power of $\log(x)$ first 
appears at either an integer or half-integer PN order.  (Note that this expands on the usage in \cite{GoldRoss10}, 
who referred only to the integral sequence in their renormalization group construction since those terms 
capture a set of UV divergences.)  To be specific, new powers of $\log(x)$ appear at integer PN orders 
$\{0,3,6,9,\ldots\}$, with the Peters-Mathews flux formally leading off this sequence.  At half-integer PN 
orders, leading logarithms occur at orders $\{3/2,9/2,15/2,\ldots\}$, which begins formally with the 1.5PN tail.

As we show in this paper, the theoretical understanding of the whole sequence of these terms is \emph{entirely} bound 
up in the Fourier spectrum of the tracefree (Newtonian) mass quadrupole moment tensor, $I_{ij}(t)$.  Let the Fourier 
amplitudes of this tensor be $I_{ij}^{(n)}$, where $n$ denotes harmonics of the Newtonian orbital frequency.  The 
leading (Peters-Mathews) quadrupole flux is proportional to the sum over $n$ of $n^6 | I_{ij}^{(n)} |^2 $.  From 
these terms we can remove factors of the reduced mass and semimajor axis to form a dimensionless function
$g(n,e_t) \coloneqq n^6 | I_{ij}^{(n)} |^2 /(16 \mu^2 a^4) $ that serves as a power spectrum for the quadrupole 
radiation.  (The function $g(n,e_t)$ is defined more completely in Sec.~\ref{sec:secII}, along with differences in 
definitions of eccentricities like $e_t$.)  The sum over $n$ of the spectrum $g(n,e_t)$ yields the well known 
Peters-Mathews enhancement function, originally called $f(e_t)$ but here called $\mathcal{R}_0(e_t)$
\be
\mathcal{R}_{0}(e_t) = \sum_{n=1}^\infty g(n,e_t) 
= \frac{1}{(1-e_t^2)^{7/2}}\bigg(1+\frac{73}{24}e_t^2 + \frac{37}{96}e_t^4\bigg) .
\ee

It turns out that a different sum over the power spectrum $g(n,e_t)$ gives rise to the eccentricity enhancement 
function $\varphi(e_t)$ for the 1.5PN tail \cite{BlanScha93} and its relative energy flux $\mathcal{R}_{3/2}(e_t)$ 
\be
\mathcal{R}_{3/2}(e_t) = 4\pi \varphi(e_t) = 4\pi \sum_{n=1}^\infty \frac{n}{2} \, g(n,e_t) .
\ee
The next sum of this type, over $(n/2)^2 \, g(n,e_t)$, produces another well known eccentricity enhancement 
function, $F(e_t)$, that is proportional to the 3PN log energy flux term $\mathcal{R}_{3L}(e_t)$ \cite{ArunETC08a}.  
Note that these three terms are the first three elements in the leading-logarithm sequence.  Furthermore, in the 
full PN analysis \cite{Blan14}, each of these fluxes only occurs at lowest order in the mass ratio.  

A new result in this paper is to show that the eccentricity dependence of the entire leading-logarithm sequence, 
which is lowest order in the mass ratio, can be understood in terms of the following sums over powers of $n/2$ that 
weight the Newtonian mass quadrupole power spectrum $g(n,e_t)$:
\begin{align}
T_k(e_t) &= \sum_{n=1}^{\infty} \Big(\frac{n}{2}\Big)^{2k}g(n,e_t) ,
\\
\Theta_k(e_t) &= \sum_{n=1}^{\infty} \Big(\frac{n}{2}\Big)^{2k+1}g(n,e_t) .
\end{align}
These sums give the eccentricity enhancement functions for integer and half-integer leading-log terms, respectively.  
We have then used BHPT calculations to verify all or part of the eccentricity dependence of the first fifteen 
elements in the leading-logarithm sequence.

However, the role of the power spectrum $g(n,e_t)$ is not confined to merely the leading-logarithm sequence.  We 
show further that the spectrum contributes two essential parts of the eccentricity dependence of each subleading 
logarithm, which are the fluxes that appear at integer PN orders $x^{3k} \log^{k-1}(x)$ and half-integer PN 
orders $x^{3k+3/2} \log^{k-1}(x)$ for $k\ge 1$.  For a given $k$, part of the integer-order subleading logarithm 
can be demonstrated to depend upon the associated leading-log enhancement function $T_k(e_t)$ and the corresponding 
($k$) sum from the added sequence
\be
\Lambda_k(e_t) = \sum_{n=1}^{\infty} \Big(\frac{n}{2}\Big)^{2k} \log\Big(\frac{n}{2}\Big) \, g(n,e_t) .
\ee
Similarly, for a given $k$, part of the half-integer-order subleading log is proportional to the leading-log enhancement 
function $\Theta_k(e_t)$ and a part is proportional to the corresponding sum in the sequence
\be
\Xi_k(e_t) = \sum_{n=1}^{\infty} \Big(\frac{n}{2}\Big)^{2k+1} \log\Big(\frac{n}{2}\Big) \, g(n,e_t) .
\ee
The remaining behavior of the subleading-logarithm terms can (in principle) be determined by BHPT calculations.  
As far as we can determine, the coefficients on $T_k(e_t)$ and $\Lambda_k(e_t)$ (or $\Theta_k(e_t)$ and 
$\Xi_k(e_t)$) within the subleading logs soak up all of the appearance of transcendental numbers.  The remaining 
eccentricity dependence in the subleading logs appears to only involve rational number coefficients.  Finally, 
we note that everything said here about energy fluxes has a mirror behavior in angular momentum fluxes.

The layout of this paper is as follows.  We first discuss in Sec.~\ref{sec:secII} the general form of the PN 
expansion for the energy and angular momentum fluxes radiated to infinity.  We then go on in that section to 
review how the Newtonian mass quadrupole moment $I_{ij}$ gives rise to the quadrupole 
radiation power spectrum $g(n,e_t)$ and how it determines not only the leading Peters-Mathews flux but also the 
1.5PN tail contribution, and the 3PN log term (the first appearance of a logarithm in the PN expansion of the flux).  
In Sec.~\ref{sec:entireLL} we use $g(n,e_t)$ to derive the sums that express the eccentricity dependence of the 
entire class of leading logarithms, giving specific examples for (9/2)L, 6L$^2$, 9L$^3$, and 12L$^4$ PN orders.  
Sec.~\ref{sec:additional6L} discusses the subleading logarithms, presenting the conjectured appearance of the 
Newtonian quadrupole spectrum in these fluxes.  We show then specific subleading-log examples at 9/2 and 6L PN 
orders, where BHPT results \cite{MunnETC19} can be combined with the PN analysis to determine the eccentricity 
dependence of the entire 6L PN term and of a lengthy power series expansion for the 9/2 PN term.

Throughout this paper we use units in which $c = G = 1$.  In discussing energy and angular momentum fluxes, 
there arise various pairs of directly comparable functions.  To distinguish a function in the angular 
momentum sector we use a tilde, e.g., $\tilde{g}(n,e_t)$, while leaving the base symbol bare, e.g., $g(n,e_t)$, 
for the energy counterpart.  This notation is in keeping with that of \cite{ArunETC08a,ArunETC08b,ArunETC09a}.  

\begin{widetext}
\section{The recurring appearance of the mass quadrupole in multiple PN contributions to the gravitational 
radiation at infinity}
\label{sec:secII}

\subsection{The post-Newtonian expansion of fluxes: general form for eccentric orbits}

We consider the post-Newtonian series for gravitational radiation at infinity.  Take two non-spinning bodies, 
a primary of mass $m_1$ and a secondary of mass $m_2$, in a bound eccentric orbit.  In the extreme-mass-ratio 
limit we will have $m_2 \ll m_1$.  We utilize a PN representation with three (dimensionless) parameters: the 
previously mentioned compactness parameter $x \coloneqq ((m_1+m_2)\Omega_\vp)^{2/3}$, the symmetric mass ratio 
$\nu=m_1m_2/(m_1+m_2)^2$, and (in modified harmonic gauge) the quasi-Keplerian time eccentricity 
$e_t$ \cite{Blan14}.  Here, $\Omega_\vp$ is the mean azimuthal orbital frequency.

In general, the parameters $x$ and $e_t$ can only be known in terms of other quantities, such as the 
energy $E$ and angular momentum $J$ of the orbit (or vice versa), as precisely as the (current) PN expansion of 
the equations of motion.  In 2004 \cite{MemmGopaScha04}, the quasi-Keplerian representation for the orbit was 
extended to 3PN order.  More recently, progress on the self-consistent center-of-mass equations has allowed explicit 
calculation of the conservative motion, and given definition to $x$, for example, to 4PN for circular orbits 
\cite{BernETC18}.  For eccentric orbits the fluxes in the dissipative sector are known as expansions in $x$ to 
3PN relative order, with half-integer terms appearing in the series starting at $x^{3/2}$ \cite{BlanScha93}.  

\subsubsection{The energy flux}

In terms of these parameters, the (orbit-averaged) energy flux is expected to have a PN expansion of the following 
form \cite{Blan14, GoldRoss10, Fuji12a, Fuji12b}:
\begin{align}
\label{eqn:energxfluxInf}
\bigg\langle \frac{dE}{dt} &\bigg\rangle_\infty = 
\frac{32}{5} \nu^2 x^5 \biggl[\mathcal{R}_0 + x\mathcal{R}_1 + x^{3/2}\mathcal{R}_{3/2}
+x^2\mathcal{R}_2 + x^{5/2}\mathcal{R}_{5/2}
+ x^3\Big(\mathcal{R}_3+\mathcal{R}_{3L}\log(x)\Big)
+ x^{7/2}\mathcal{R}_{7/2}
\notag\\&+ x^4\Big(\mathcal{R}_4+\mathcal{R}_{4L}\log(x)\Big)
+x^{9/2}\Bigl(\mathcal{R}_{9/2}+\log(x)\mathcal{R}_{9/2L}\Bigr)
+x^5\Bigl(\mathcal{R}_5
+\log(x)\mathcal{R}_{5L}\Bigr)
+x^{11/2}\Bigl(\mathcal{R}_{11/2}+\log(x)\mathcal{R}_{11/2L}\Bigr)
\notag\\& 
+ x^6\Bigl(\mathcal{R}_6 + \log(x)\mathcal{R}_{6L}
+ \log^2(x)\mathcal{R}_{6L2} \Bigr)
+x^{13/2}\Bigl(\mathcal{R}_{13/2}+\log(x)\mathcal{R}_{13/2L}\Bigr)
+\cdots
\biggr],
\end{align}
\end{widetext}
where each $\mathcal{R}_i$ is (in general) a function of $e_t$ and $\nu$.  Since we are principally interested in 
the overlap between PN theory and BHPT, at first order in the mass ratio each 
$\mathcal{R}_i$ can be evaluated at $\nu=0$.  In this paper, these functions will thus simply be taken as depending 
on $e_t$ alone: $\mathcal{R}_i = \mathcal{R}_i(e_t)$.  Each such function is known to diverge as $e_t \rightarrow 1$.  
Because the Peters-Mathews function \cite{PeteMath63} $\mathcal{R}_0(e_t)$ has the limit $\mathcal{R}_0 = 1$ as
$e_t\rightarrow 0$, the prefactor in the above expansion is simply the Newtonian (quadrupole) circular-orbit energy 
flux, which can be further reduced to $(32/5) \nu^2 (m_1 + m_2)^5/a^5$ in terms of the semimajor axis $a$ in the 
Newtonian limit. 

In PN derivations, a distinction is often made between instantaneous and hereditary contributions to the flux 
that alternately or simultaneously appear at different PN orders.  The hereditary terms depend on the entire 
history of the system (see, for instance, \cite{Blan14}).  However, when BHPT is applied to wide orbits, the flux 
terms (at lowest order in the mass ratio) that emerge in a subsequent PN expansion are a sum of instantaneous and 
hereditary parts, as the method does not generally distinguish between the two (though see 
Sec.~\ref{sec:MoreGeneral} for more discussion and cases where some distinction is possible).  With this in mind, 
in this paper we simply use $\mathcal{R}_i(e_t)$ at each order in $x$ to represent the sum of both 
instantaneous and hereditary contributions.

One route often taken in BHPT calculations is to work in the frequency domain and evaluate the 
self-force, at lowest order in the mass ratio, using a geodesic in the background spacetime.  For a non-spinning 
primary, geodesics are computed in Schwarzschild spacetime using (typically) Schwarzschild coordinates.  Bound 
eccentric orbits are frequently described by the relativistic Darwin \cite{Darw59,Darw61} eccentricity $e$ and 
(dimensionless) semi-latus rectum $p$.  When this approach is applied to wide orbits, a PN expansion can be derived, 
using typically the alternate compactness parameter $y \coloneqq (m_1 \Omega_\vp)^{2/3}$.  Expansions in this 
form were made in an earlier paper \cite{ForsEvanHopp16} in this series (and used \cite{MunnETC19} in a companion
paper).  When $y$ and $e$ are used, the PN expansion of the energy flux is similar in form to 
\eqref{eqn:energxfluxInf} except now the flux functions $\mathcal{L}_i(e)$ depend on Darwin $e$.  While the 
parameters $(y, e)$ can be expressed in terms of $(x, e_t)$ through expansions that begin with
$y = x (1 - 2 \nu/3 + \mathcal{O}(\nu^2) )$ and $e = e_t (1 + 3 x + \mathcal{O}(\nu,e_t,x^2))$, it is clear that 
in general $\mathcal{L}_i(e) \ne \mathcal{R}_i(e_t)$.  Exceptions are when order $i$ terms emerge purely from 
Newtonian quantities.  For most of the present paper, we opt to use $(x,e_t)$ and the standard PN expansion in 
the form \eqref{eqn:energxfluxInf}.  However, the $\mathcal{L}_i(e)$ notation will reappear in 
Sec.~\ref{sec:additional6L}, when our PN derivations are combined with BHPT numerical 
results to extract the full $\mathcal{L}_{6L}$ term.

As mentioned in the Introduction, a leading logarithm term is defined as one in which a new higher power of 
$\log(x)$ first appears, at both integer and half-integer PN orders.  New powers of $\log(x)$ 
appear at integer PN orders $\{0,3,6,9,\ldots\}$, which includes the Peters-Mathews term that has $\log^0(x)$.  
New powers of log appear at half-integer PN orders $\{3/2,9/2,15/2,\ldots\}$.  Thus, the leading logarithm 
portion of the series \eqref{eqn:energxfluxInf} has the form
\begin{widetext}
\begin{align}
\label{eqn:energyfluxLL}
\left\langle \frac{dE}{dt} \right\rangle_\infty^{\text{LL}} 
= \frac{32}{5} \nu^2 x^5 \biggl[\mathcal{R}_0 
&+ x^{3/2}\mathcal{R}_{3/2} 
+ x^3\log(x)\mathcal{R}_{3L} +x^{9/2}\log(x)\mathcal{R}_{9/2L} + x^6\log^2(x)\mathcal{R}_{6L2} 
\notag 
\\
&+x^{15/2}\log^2(x) \mathcal{R}_{15/2L2} + x^9 \log^3(x) \mathcal{R}_{9L3}+\cdots \biggr] .
\end{align}
One of the principal results of this paper, as we will show in Sec.~\ref{sec:entireLL}, is that the analytic
eccentricity dependence of this entire infinite sequence can be determined in straightforward fashion
using the Newtonian mass quadrupole.  Integer-order terms will in fact yield closed-form expressions, while
half-integer-order terms will yield infinite convergent expansions in $e_t$ that can be rapidly generated 
to arbitrary order.  Because of the origin of these terms, a side effect is that we have
$\mathcal{R}_{i}^{\rm LL}(e_t) = \mathcal{L}_{i}^{\rm LL}(e)$ for every term in \eqref{eqn:energyfluxLL}.

\subsubsection{The angular momentum flux}

The angular momentum flux has a similar expected PN expansion
\begin{align}
\label{eqn:angMomFluxInf}
\bigg\langle \frac{dL}{dt} &\bigg\rangle_\infty =  
\frac{32}{5} \nu^2 (m_1 + m_2) x^{7/2}
\biggl[\mathcal{Z}_0 + x\mathcal{Z}_1 + x^{3/2}\mathcal{Z}_{3/2}
+x^2\mathcal{Z}_2 + x^{5/2}\mathcal{Z}_{5/2}
+ x^3\Big(\mathcal{Z}_3+\mathcal{Z}_{3L}\log(x)\Big)
+ x^{7/2}\mathcal{Z}_{7/2}
\notag\\&+ x^4\Big(\mathcal{Z}_4+\mathcal{Z}_{4L}\log(x)\Big)
+x^{9/2}\Bigl(\mathcal{Z}_{9/2}+\log(x)\mathcal{Z}_{9/2L}\Bigr)
+x^5\Bigl(\mathcal{Z}_5
+\log(x)\mathcal{Z}_{5L}\Bigr)
+x^{11/2}\Bigl(\mathcal{Z}_{11/2}+\log(x)\mathcal{Z}_{11/2L}\Bigr)
\notag\\& 
+ x^6\Bigl(\mathcal{Z}_6 + \log(x)\mathcal{Z}_{6L}
+ \log^2(x)\mathcal{Z}_{6L2} \Bigr)
+x^{13/2}\Bigl(\mathcal{Z}_{13/2}+\log(x)\mathcal{Z}_{13/2L}\Bigr)
+\cdots
\biggr] , 
\end{align}
\end{widetext}
where again each $\mathcal{Z}_i$ is generally a function of both $e_t$ and $\nu$.  At first order in the mass 
ratio we will simply take $\mathcal{Z}_i = \mathcal{Z}_i(e_t)$, and these terms are meant to combine both 
instantaneous and hereditary contributions.  The leading logarithm series in this case has the same form as 
\eqref{eqn:energyfluxLL} but with the substitutions $(32/5)\nu^2 x^5 \rightarrow (32/5)\nu^2 (m_1+m_2) x^{7/2}$ 
and $\mathcal{R} \rightarrow \mathcal{Z}$. 
 
In both fluxes, the eccentricity functions at any given PN order can be derived from time derivatives (and 
potentially integrals) of mass and current multipole moments of the system.  In general, higher PN order requires 
higher multipole moments, and their derivatives and PN corrections.  The lowest-order multipole moment that appears 
in these fluxes is the tracefree part of the Newtonian (0PN) mass quadrupole moment, $I_{ij}$, found through 
calculation on a Newtonian orbit.  It is from this tensor that $\mathcal{R}_0$ \cite{PeteMath63} and $\mathcal{Z}_0$ 
\cite{Pete64} were first derived.  At 1PN in the fluxes, the 0PN mass octupole and current quadrupole moments 
appear, as well as the 1PN correction to the mass quadrupole (which entails quadrupole moment calculation on a 
precessing 1PN orbit) \cite{Blan14}.  In turn, at 2PN in the fluxes, the 0PN mass hexadecapole and current octupole 
appear, as well as 1PN corrections to the mass octupole and current quadrupole and 2PN correction to the mass 
quadrupole.  

In this paper, we determine PN flux content that is generated exclusively by the 0PN mass quadrupole.  
However, it is not difficult to see that extending the procedures outlined here to higher multipole moments 
and their PN corrections will yield additional analytic pieces of comparable depth in other terms in the PN 
expansion.  Such an exploration at the 1PN correction level has in fact been successful, and results will be 
reported in a subsequent paper.  

\subsection{The quadrupole moment and the Kepler problem}

We briefly review the calculation of the Newtonian quadrupole to derive functions that are essential for the 
rest of the paper.  The analysis starts with the Kepler motion problem for bound, elliptical orbits and uses the 
Fourier series expansion for its time dependence.  The masses are constrained to the $x-y$ plane and the relative 
motion is described in terms of polar coordinates $r=r(t)$ and $\vp=\vp(t)$ for the separation and azimuthal 
angle, respectively.  Because our preferred time eccentricity $e_t$ reduces to the usual Keplerian eccentricity 
at 0PN order, $r$ and $\vp$ can simply be given by
\be
\label{eqn:rphiPhidot}
r=\frac{a(1-e_t^2)}{1+e_t\cos\vp}, \qq \dot\vp^2=\frac{a \left(1-e_t^2\right) M}{r^4},
\ee
where $M={m_1}+{m_2}$.

Summing over the two bodies, the gravitational wave fluxes will be obtained from the components of the 
tracefree mass quadrupole tensor
\begin{gather}
I_{xx} = \mu r^2 \cos^2\vp - \mu r^2/3, \notag \\ 
I_{xy} =I_{yx} = \mu r^2 \sin\vp\cos\vp, \notag \\
I_{yy} = \mu r^2 \sin^2\vp - \mu r^2/3, \notag \\
I_{zz} = -\mu r^2/3 . \label{eqn:rphiMom}
\end{gather}
Here $\mu=m_1 m_2/M$ is the reduced mass of the system and $r$ and $\vp$ are evaluated as functions of some curve 
parameter.  A convenient such choice is the eccentric anomaly $u=\arccos((a-r)/ae_t)$, which yields
\begin{align}
I_{xx} &= \frac{1}{6} \mu a^2  \left(1+5 e_t^2-8 e_t \cos u-\left(e_t^2-3\right) \cos 2u \right), \notag \\ \notag
& \quad I_{xy} =I_{yx} = \mu a^2 \sqrt{1-e_t^2} (\cos u-e_t) \sin u, \\ \notag
I_{yy} &= \frac{1}{6} \mu a^2  \left(1-4 e_t^2+4 e_t \cos u+\left(2 e_t^2-3\right) \cos 2u \right), \\ 
&\qquad \qquad I_{zz} = -\frac{1}{3} \mu a^2  (e_t \cos u-1)^2. \label{eqn:aeMom}
\end{align}

Since the tensor components \eqref{eqn:aeMom} are all periodic functions of $u$ (or $t$), each can be written as 
a Fourier series.  Following the discussion of Arun \cite{ArunETC08a}, we write
\begin{equation}
\label{eqn:fourier}
I_{ij} = \sum_{n=-\infty}^{\infty} I_{ij}^{(n)} e^{inl}, 
\end{equation}
where $ I_{ij}^{(n)}$ is the $n$th Fourier component of $I_{ij}$, and $l$ is the mean anomaly of the motion
\begin{align}
l= u-e_t\sin u=\frac{2\pi}{T_r}(t-t_P) = \Omega_r(t-t_P) .
\end{align}
Here $T_r$ is the radial libration period, $\Omega_r$ the radial angular frequency (equal to $\Omega_\vp$ in the 
Newtonian limit), and $t_P$ the time of periastron crossing.  The Fourier components are derived from
\begin{equation}
I_{ij}^{(n)} = \frac{1}{2\pi} \int_{0}^{2\pi}I_{ij}(u(l)) \, e^{-inl} \, dl .
\end{equation}

The Fourier series coefficient integrals are taken over mean anomaly (or time) while the quadrupole moment components 
are sinusoidal functions of $u$.  We can evaluate these integrals in several ways but the easiest is to write them 
in terms of $u$ 
\begin{equation}
\label{eqn:FourIntU}
I_{ij}^{(n)} = \frac{1}{2\pi} \int_{0}^{2\pi}I_{ij} e^{-in(u-e_t \sin u)}(1-e_t \cos u) du.
\end{equation}
Once the various circular functions have been recast as complex exponentials, \eqref{eqn:FourIntU} will reduce to 
a sum of Bessel integrals \cite{DLMF} of the form
\begin{equation}
\frac{1}{2\pi} \int_{0}^{2\pi} e^{-ipu + i x \sin u} du = J_p(x).
\end{equation}
Then these are simplified using Bessel function identities (see \cite{PeteMath63,BlanScha93} for similar 
derivation) to obtain
\begin{align}
\label{eqn:fourierquad}
\begin{autobreak}
\MoveEqLeft
I_{xx}^{(n)} = 2 \mu a^2 \bigg[
\frac{e_t^2-3}{3 n^2 e_t^2}J_n(n e_t) + \frac{1 - e_t^2}{n e_t} J_n'(n e_t)\bigg]
\end{autobreak}, \notag \\

\begin{autobreak}
\MoveEqLeft
I_{xy}^{(n)} =I_{yx}^{(n)} =\mu a^2\bigg(\frac{2 i \sqrt{1-e_t^2}}{n e_t}\bigg) \bigg[-\frac{1-e_t^2}{e_t}J_n(n e_t)
+\frac{1}{n} J_n'(n e_t)  \bigg],
\end{autobreak} \notag \\

\begin{autobreak}
\MoveEqLeft 
I_{yy}^{(n)} = 2 \mu a^2\bigg[\frac{3 - 2 e_t^2}{3 n^2 e_t^2} J_n(n e_t)
- \frac{1-e_t^2}{n e_t} J_n'(n e_t) \bigg]
\end{autobreak},\notag \\ 

\begin{autobreak}
\MoveEqLeft
I_{zz}^{(n)} =\mu a^2\frac{2 J_n(n e_t)}{3 n^2}
\end{autobreak}.
\end{align}

\subsection{The power spectra $g(n,e_t)$ and $\tilde{g}(n,e_t)$ and the Peters-Mathews enhancement functions}

With these expressions in hand, the Newtonian-order energy and angular momentum fluxes can be found using the 
classic formulas 
\begin{align}
\label{eqn:I0}
\Big<\frac{dE}{dt}\Big>_N &= \frac{1}{5}\Big<\dddot{I}_{ij}\dddot{I}_{ij}\Big>, \\
\label{eqn:H0}
\Big<\frac{dL}{dt}\Big>_N =&\frac{2}{5}\epsilon_{ijk} \hat{L}_i \Big<\ddot{I}_{ja}\dddot{I}_{ka}\Big> , 
\end{align}
where angled brackets denote the time average over an orbital period and $\hat{L}_i$ is the unit vector in the 
angular momentum direction, which here is $\hat{L}_i = (0,0,1)$.

\subsubsection{The function $g(n,e_t)$ and the spectral content of the Newtonian quadrupole energy flux}

For the energy flux, a Fourier decomposition of \eqref{eqn:I0} can be found from a double application of the sum 
\eqref{eqn:fourier}, giving
\begin{align}
\begin{autobreak}
\MoveEqLeft
\Big<\frac{dE}{dt}\Big>_N 
= \frac{1}{5} \bigg\langle \sum_{n_1=-\infty}^{\infty} \sum_{n_2=-\infty}^{\infty} (in_1\Omega_r)^3 (in_2\Omega_r)^3 
\qquad \qquad \qquad \qquad \times I_{ij}^{(n_1)}  I_{ij}^{(n_2)} e^{i(n_1+n_2)l}\bigg\rangle
\end{autobreak}\notag \\
&\qquad = \frac{2}{5} (\Omega_r)^6  \sum_{n=1}^{\infty} n^6 I_{ij}^{(n)}  I_{ij}^{(n)*} .
\end{align}
The final equality follows from the time average giving $\delta_{n_1,-n_2}$ and, because $I_{ij}(t)$ is real, 
from the crossing relations $I_{ij}^{(-n)} = I_{ij}^{(n)*}$ on the Fourier coefficients.   

A dimensionless portion of the energy flux can be isolated and normalized by removing a factor of 
$16\mu^2 a^4$ (which generalizes to $16 \mu^2 M^4/x^4$ beyond Newtonian order), leading to 
\begin{align}
&\quad \Big<\frac{dE}{dt}\Big>_N= \frac{32}{5}(\Omega_r)^6 \mu^2 a^4 \sum_{n=1}^{\infty} g(n,e_t), \\
&\qquad \qquad   g(n,e_t) \coloneqq \frac{1}{16 \mu^2 a^4} n^6 | I_{ij}^{(n)} |^2 . 
\label{eqn:gdimen}
\end{align}
As is obvious from the expression above, the dimensionless function $g(n,e_t)$ (first derived in \cite{PeteMath63} 
and then corrected in \cite{BlanScha93}) represents the (relative) power radiated in the $n$-th harmonic of the 
orbital frequency (i.e., the power spectrum).  Combining \eqref{eqn:fourierquad} and \eqref{eqn:gdimen}, this 
function is found to be
\begin{align}
g(n,e_t)&= \frac{n^2}{2} \bigg\{ \left[-\frac{4}{e_t^3}-3 e_t +
\frac{7}{e_t}\right] n J_n(n e_t) J_n'(n e_t) \, \notag \\
+\bigg[\frac{1}{e_t^4}&-\frac{1}{e_t^2}+
\left(\frac{1}{e_t^4}-e_t^2-\frac{3}{e_t^2}+3\right) n^2+\frac{1}{3}\bigg] 
 J_n(ne_t)^2
\notag \\ &\mkern-25mu  + \left[\left(e_t^2+\frac{1}{e_t^2}-2\right) n^2+\frac{1}{e_t^2}-1\right] 
J_n'(n e_t)^2 
 \bigg\} . 
\label{eqn:gfunc}
\end{align}

The total power is the sum of $g(n,e_t)$ over all harmonics, which once computed yields the first example of an
eccentricity enhancement function (so named because eccentric orbits have enhanced flux 
relative to a circular orbit of the same $a$ or orbital frequency $\Omega_\vp$).  Straightforwardly summing this 
function yields an infinite series in $e_t$
\be
\mathcal{R}_0 = \sum_{n=1}^{\infty} g(n,e_t) = 1 + \frac{157}{24}e_t^2 + \frac{605}{32}e_t^4 + 
\frac{3815}{96}e_t^6 +\cdots .
\ee
A cleaner result is found by introducing the known eccentricity singular factor $(1-e_t^2)^{-7/2}$ and 
resumming the series to find a closed form expression
\be
\label{eqn:PMf}
\mathcal{R}_0(e_t) = \frac{1}{(1-e_t^2)^{7/2}}\bigg(1+\frac{73}{24}e_t^2 + \frac{37}{96}e_t^4\bigg) ,
\ee
which is the classic result from Peters and Mathews \cite{PeteMath63}.

\subsubsection{The function $\tilde{g}(n,e_t)$ and the spectral content of the Newtonian quadrupole angular 
momentum flux}

Similarly, we plug \eqref{eqn:fourier} into \eqref{eqn:H0} and find
\begin{align}
\begin{autobreak}
\MoveEqLeft
\Big<\frac{dL}{dt}\Big>_N = \frac{2}{5} \epsilon_{ijk} \hat{L}_i \bigg\langle \sum_{n_1=-\infty}^{\infty} \sum_{n_2=-\infty}^{\infty} (in_1\Omega_r)^2 (in_2\Omega_r)^3 
\qquad \qquad \qquad \qquad \times I_{ja}^{(n_1)}  I_{ka}^{(n_2)} e^{i(n_1+n_2)l}\bigg\rangle
\end{autobreak}\notag \\

&\qquad = -\frac{4}{5} (\Omega_r)^5 i \epsilon_{ijk} \hat{L}_i \sum_{n=1}^{\infty} n^5 I_{ja}^{(n)}  I_{ka}^{(n)*}  \notag \\
&\qquad \quad = \frac{32}{5}(\Omega_r)^5 \mu^2 a^4 \sum_{n=1}^{\infty} \tilde{g}(n,e_t) ,
\end{align}
where $\tilde{g}(n,e_t)$ is given by
\begin{align}
\tilde{g}(n,e_t) \coloneqq \frac{-i}{8 \mu^2 a^4} \epsilon_{ijk} \hat{L}_i  n^5 I_{ja}^{(n)}  I_{ka}^{(n)*}.
\end{align}
The dimensionless function $\tilde{g}(n,e_t)$ mirrors its energy flux counterpart and is found to be
\begin{align}
\label{eqn:gtilde}
\tilde{g}(n,e_t) &=
\sqrt{1-e^2} \bigg\{\left[-\frac{2}{e_t^2}+2\right] n^2 J_n'(n e)^2 \notag \\
& + \frac{n}{e_t^3} \bigg[ 2-e_t^2 + 2 n^2 (1- e_t^2)^2 \bigg] J_n(ne_t) J_n'(ne_t) \notag \\
& +\left[-\frac{2}{e_t^4}+\frac{3}{e_t^2}-1\right] n^2 J_n(ne_t)^2 \bigg\} ,
\end{align}
which represents the (relative) power spectrum for angular momentum radiated per harmonic of the orbital 
frequency.

The sum of $\tilde{g}(n,e_t)$ over all $n$ can be used to obtain the Newtonian quadrupole angular momentum 
enhancement function, which was originally derived by Peters \cite{Pete64}.  Pulling out the eccentricity 
singular factor $(1-e_t^2)^{-2}$ (in this case) leads to
\be
\mathcal{Z}_0(e_t) = \sum_{n=1}^{\infty} \tilde{g}(n,e_t) 
= \frac{1}{(1-e_t^2)^{2}}\bigg(1+\frac{7}{8}e_t^2\bigg) .
\ee

\subsubsection{Discussion}

The Newtonian quadrupole power spectra, $g(n,e_t)$ and $\tilde{g}(n,e_t)$, will be shown in this paper to be the 
exclusive factors that determine the eccentricity dependence of all the higher-PN leading-log terms.  In summing 
these functions directly, particular eccentricity singular factors appeared in $\mathcal{R}_0$ and $\mathcal{Z}_0$, 
revealing the remaining part of these enhancement functions to be polynomials (which are of course finite as 
$e_t \rightarrow 1$), giving the expressions closed forms.  These two eccentricity singular factors were identified 
in the original derivations~\cite{PeteMath63,Pete64}.  As shown by more recent asymptotic analysis in 
\cite{ForsEvanHopp16,LoutYune17,MunnETC19}, enhancement functions at other PN orders have predictable singular 
factors.  Specifically, we can see in those results that sums of the form $\sum n^k g(n,e_t)$ will have the singular 
dependence $1/(1-e_t^2)^{(7/2+3k/2)}$ while those of the type $\sum n^k \tilde{g}(n,e_t)$ will carry a factor 
of $1/(1-e_t^2)^{(2+3k/2)}$.  These factors will be essential to extracting from $g$ and $\tilde{g}$ new 
closed-form expressions for the higher-PN order leading-log enhancement functions.

\subsection{Other enhancement functions already known to depend only upon $g(n,e_t)$ and $\tilde{g}(n,e_t)$}

Although the original application of $g(n,e_t)$ and $\tilde{g}(n,e_t)$ (summing them directly) was to derive 
the Newtonian (0PN) order fluxes, these functions were each later found to determine three additional 
enhancement functions.  

\subsubsection{The 1.5PN tail functions $\vp(e_t)$ and $\tilde{\vp}(e_t)$}

The first of these is the 1.5PN energy enhancement function $\vp(e_t)$ (proportional to $\mathcal{R}_{3/2}$), 
which was found in \cite{BlanScha93} to be the lowest-order tail correction to the Newtonian-order flux.  Blanchet 
and Schafer evaluated the relevant sum numerically and plotted the enhancement function.  Later, Arun 
et al.~\cite{ArunETC08a} provided the first two coefficients of a power series for $\vp(e_t)$ and then 
\cite{ForsEvanHopp16} used the Bessel representation \eqref{eqn:gfunc} to compute analytic coefficients to 
arbitrary powers of $e_t^2$.  By combining that expansion with the expected eccentricity singular function, 
the resummed power series expansion was shown \cite{ForsEvanHopp16} to be convergent for all $e_t$.
The required sum over $g(n,e_t)$ and leading part of the expansion are
\begin{align}
\label{eqn:vphiExpand}
\varphi(e_t) &= \sum_{n=1}^{\infty} \Big(\frac{n}{2}\Big)g(n,e_t) \notag \\
&=\frac{1}{(1-e_t^2)^5} \, 
\bigg(1+\frac{1375}{192} e_t^2+\frac{3935}{768} e_t^4 
+\frac{10007}{36864} e_t^6 \notag \\
&+\frac{2321}{884736} e_t^8  
-\frac{237857}{353894400} e_t^{10}
+\cdots \bigg) ,
\end{align}
(which corrects a sign error in \cite{ForsEvanHopp16}) on the $e_t^{10}$ term).  Like most enhancement functions, 
$\vp(e_t)$ is defined such that its circular orbit limit is unity.  The full (relative) energy flux term at 1.5PN 
order is 
\be 
\mathcal{R}_{3/2}(e_t) = 4 \pi \vp(e_t). 
\ee

Thus, a series proportional to the 1.5PN tail term emerges directly from a sum over $n$ of the 
$g(n,e_t)$ amplitudes multiplied by the factor $n/2$.  Unfortunately, \eqref{eqn:vphiExpand} is an infinite 
series, with $\vp$ not expected \cite{BlanScha93} to have a closed form representation.  However, by multiplying
the sum in \eqref{eqn:vphiExpand} by $(1-e_t^2)^5$ and expanding in a MacLaurin series in $e_t$, the coefficients 
each involve a finite sum in $n$ and are easily found to hundreds of orders in $e_t$ in a matter of seconds 
using \textsc{Mathematica}.  The eccentricity singular factor exponent was chosen to be $-5$ ($k=1$) in accordance 
with the earlier discussion. 

The 1.5PN angular momentum enhancement function follows similarly and can be found in \cite{ArunETC09a} (though 
without explicit mention of $\tilde{g}(n,e_t)$)
\begin{align}
\label{eqn:vphiTildExpand}
\tilde{\varphi}(e_t) &= \sum_{n=1}^{\infty} \Big(\frac{n}{2}\Big)\tilde{g}(n,e_t) \notag \\
&=\frac{1}{(1-e_t^2)^{7/2}} \, 
\bigg(1+\frac{97}{32} e_t^2+\frac{49}{128} e_t^4 
-\frac{49}{18432} e_t^6 \notag \\
&-\frac{109}{147456} e_t^8  
-\frac{2567}{58982400} e_t^{10}
+\cdots \bigg) ,
\end{align}
with its own eccentricity singular factor, which leaves an infinite series that is convergent for all $e_t$.  In 
fact, all the summations over $g(n,e_t)$ considered in this paper can be translated from giving energy flux terms 
to giving angular momentum flux terms by making the simple substitution $g \rightarrow \tilde{g}$.  Hence, for the 
rest of the paper, we focus almost exclusively on the energy flux contributions, with it being obvious how the 
corresponding angular momentum flux terms are determined.  Our full compilation of all of these enhancement 
functions can be found at \cite{BHPTK18}.

\subsubsection{The 3PN functions $F(e_t)$ and $\chi(e_t)$}

As Arun et al.~\cite{ArunETC08a,ArunETC09a} showed, the Newtonian mass quadrupole makes an
appearance again at 3PN relative order in the flux in two additional enhancement functions:
\begin{align}
F(e_t) &= \sum_{n=1}^{\infty} \Big(\frac{n^2}{4}\Big) \, g(n,e_t), \\
\chi(e_t) &= \sum_{n=1}^{\infty} \Big(\frac{n^2}{4}\Big) \, \log\Big(\frac{n}{2}\Big) \, g(n,e_t) .
\end{align}
Because of the even power of $n$ in its summation, $F(e_t)$ turns out to have its own closed-form expression  
\begin{align}
F(e_t) &= \frac{1}{(1-e_t^2)^{13/2}} 
\bigg( 1+ \frac{85}{6} e_t^2 + \frac{5171}{192} e_t^4 \notag \\
&\qq \q +\frac{1751}{192} e_t^6 + \frac{297}{1024} e_t^8\bigg) .
\label{eqn:capFe}
\end{align}
This result follows from being able to convert the sum over Fourier amplitudes to an integral over time (time 
average) in the time domain (i.e., application of Parseval's theorem).  The result is proportional to the integral 
of the square of the fourth time 
derivative, $\langle\prescript{(4)}{}{I}_{ij}\prescript{(4)}{}{I}_{ij}\rangle$ \cite{ArunETC08a}, which once 
integrated becomes \eqref{eqn:capFe}.  Here prescripts indicate time derivatives of moments, e.g., 
$\prescript{(2)}{}I_{ij}(t) = d^2 I_{ij}(t)/dt^2$, which should not be confused with Fourier coefficients, such as 
$I_{ij}^{(n)}$.

The $\log(n/2)$ factor in the sum for the enhancement function $\chi(e_t)$ all but ensures that it will not have 
a closed form.  (While $\chi(e_t)$ is referred to as an enhancement function, it is a rare case of one that 
vanishes as $e_t \rightarrow 0$ \cite{ArunETC08a}.)  As with $\vp(e_t)$, the best option is to isolate a 
convergent series in $e_t$ that can be calculated to arbitrary order as needed.  As shown in \cite{ForsEvanHopp16}, 
that process involves identifying and pulling out a particular term that is both logarithmically and power-law 
divergent and then determining the remaining expansion
\begin{align}
\begin{autobreak}
\label{eqn:chi}
\MoveEqLeft
\chi (e_t) = - \frac{3}{2} F(e_t) \log(1 - e_t^2) +
\frac{1}{(1-e_t^2)^{13/2}} \, \Bigg[ 
\bigg(-\frac{3}{2} - \frac{77}{3} \log(2) + \frac{6561}{256} \log(3) \bigg) e_t^2 
+
\bigg(-22 + \frac{34855}{64} \log(2) - \frac{295245}{1024} \log(3) \bigg) e_t^4 
+
\bigg(-\frac{6595}{128} - \frac{1167467}{192} \log(2) + 
\frac{24247269}{16384} \log(3) 
+ \frac{244140625}{147456} \log(5) \bigg) e_t^6 
+ \cdots \Bigg] .
\end{autobreak}
\end{align}
The infinite series in square braces then turns out to be convergent for all $e_t$.  Interestingly, the function 
$F(e_t)$ itself appears in a term with logarithmic divergence as $e_t\rightarrow 1$, and thus plays an essential 
role in the expansion of $\chi(e_t)$.  This makes $\chi(e_t)$ possess not only the expected eccentricity singular 
factor for a 3PN enhancement function, $(1 - e_t^2)^{-13/2}$, but also a separate logarithmic/power-law 
divergence.  This fact will be important in Sec.~\ref{sec:additional6L} where we study the structure of the 
subleading logarithms (defined in the Introduction).  What we show is that each subleading logarithm is intimately 
connected to its associated leading logarithms (e.g., at 6PN the subleading term $\mathcal{R}_{6L}$ bears some 
functional connection to the $\mathcal{R}_{6L2}$ leading log).  

The first such connection between the two sequences occurs at 3PN order.  The following sum, of 3PN 
log (a leading log) and 3PN (a subleading log), is equal to the full 3PN (relative) flux \cite{ArunETC08b} at lowest 
order in the mass ratio
\begin{widetext}
\begin{align}
\label{eqn:3PNE}
\mathcal{R}_3 +\mathcal{R}_{3L}\log x &= 
\frac{1}{(1-e_t^2)^{13/2}}\biggl[
\frac{2193295679}{9979200}+\frac{20506331429}{19958400} e_t^2-\frac{3611354071}{13305600} e_t^4
+\frac{4786812253}{26611200} e_t^6+\frac{21505140101}{141926400} e_t^8 \notag\\
&-\frac{8977637}{11354112} e_t^{10}
+\sqrt{1-e_t^2}\left(-\frac{14047483}{151200}+\frac{36863231}{100800} e_t^2
+\frac{759524951}{403200} e_t^4 + \frac{1399661203}{2419200} e_t^6 + \frac{185}{48} e_t^8
 \right) \biggr]\notag\\
&+\left(
\frac{16}{3}\pi^2-\frac{1712}{105}\gamma_\text{E}-\frac{116761}{3675}
\right)F(e_t)
- \frac{856}{105}\log \left(\left[\frac{8 \left(1-e_t^2\right)}{1
+\sqrt{1-e_t^2}}\right]^2 x\right)F(e_t)
-\frac{1712}{105}\chi(e_t) .
\end{align}
\end{widetext}
This expression shows a distinctive manner in which the functions $\chi(e_t)$ and $F(e_t)$ combine in the net 3PN 
flux.  Both functions are known to contribute \cite{ArunETC08a} to the tail-of-tail and $\text{tail}^2$ hereditary 
pieces.  These two functions also are associated with all of the transcendental numbers in the flux.  Clearly, one 
appearance of the function $F(e_t)$ above can be seen to gather all of the obvious transcendental numbers, like 
$\pi^2$ and the Euler-Mascheroni constant $\gamma_\text{E}$.  However, the expansion of $\chi(e_t)$ \eqref{eqn:chi} 
reveals added transcendentals.  The gathering of all the transcendentals on $F(e_t)$ and $\chi(e_t)$ at 3PN has an 
analogue at higher PN orders that will be exploited in Sec.~\ref{sec:additional6L}.

Though it is not apparent in \eqref{eqn:3PNE}, $F$ appears also in the instantaneous part \cite{ArunETC08a}.  
Upon examining \eqref{eqn:3PNE} more closely, we see that every part of the total 3PN flux has a closed-form 
representation except the $\chi(e_t)$ term, which is an infinite series.  In addition, $F(e_t)$ multiplies an 
obvious divergent logarithm of $1 - e_t^2$, but the same term with a different coefficient appears in the 
expansion of $\chi(e_t)$.  Finally, what is most significant for the discussion in this section is that $F(e_t)$ 
is proportional to the $\log x$ term, which means that
\be
\mathcal{R}_{3L}(e_t)=-\frac{856}{105}F(e_t) .
\ee
So, except for a rational numerical factor that gives the circular orbit limit, a sum over the Newtonian 
mass quadrupole Fourier spectrum $g(n,e_t)$ gives the entire $\mathcal{R}_{3L}$ flux function, which is a 
closed-form expression.

All of the discussion here pertains equally well to the full 3PN angular momentum flux and analogous enhancement 
functions $\tilde{F}(e_t)$ and $\tilde{\chi}(e_t)$ obtained from $\tilde{g}(n,e_t)$ \cite{ArunETC09a}.

\section{Obtaining the entire leading logarithm sequence from the mass quadrupole power spectra $g(n,e_t)$ 
and $\tilde{g}(n,e_t)$}
\label{sec:entireLL}

As the review in the last section has shown, the eccentric-orbit Newtonian mass quadrupole spectrum $g(n,e_t)$ 
is solely responsible for determining the first three leading-log eccentricity functions, $\mathcal{R}_{0}$, 
$\mathcal{R}_{3/2}$, and $\mathcal{R}_{3L}$.  These flux terms emerged from sums over $g(n,e_t)$ times 
factors of $n/2$ to the first three integer powers.  In this paper, we show that this progression continues to 
higher PN order, with additional leading-log terms being determined exclusively by sums over $g(n,e_t)$ times 
increasing powers of $n/2$.  The progression splits into two infinite sequences for even and odd powers of $n/2$, 
which correspond to fluxes at integer and half-integer powers of $x$, respectively.

\subsection{All leading-log enhancement functions at integer powers of $x$ have closed-form expressions}

As we briefly touched on in the Introduction, we first consider all sums over the product of the Newtonian mass 
quadrupole spectrum $g(n,e_t)$ and even powers of $n/2$
\be
\label{eqn:capTk}
T_k(e_t)=\sum_{n=1}^{\infty} \Big(\frac{n}{2}\Big)^{2k}g(n,e_t) ,
\ee
where $k \ge 0$ is an integer.  Under this definition, $T_0(e_t) = \mathcal{R}_0(e_t)$ and $T_1(e_t) = F(e_t)$.
With even powers of $n$, every one of these sums can be converted to the time domain and shown to be 
proportional to an integral (time average) of products of time derivatives of $I_{ij}(t)$
\be
\label{eqn:TDmassquads}
\Big<\prescript{(k+3)}{}I_{ij}(t) \, \prescript{(k+3)}{}I_{ij}(t) \Big> .
\ee
If instead we view this in reverse, and convert \eqref{eqn:TDmassquads} to the frequency domain, then each time 
derivative carries with it a factor of $\Omega_r = x^{3/2}/M+\mathcal{O}(x^{5/2})$.  Since the Newtonian relative
order flux \eqref{eqn:I0} itself carries a factor of $\Omega_r^6$ (i.e., \eqref{eqn:gdimen}), each $T_k$ will be 
a $(3k)$PN order quantity.  Furthermore, it can be shown that the resulting expression will be singular 
as $e_t\rightarrow 1$ and that the singular dependence is captured for each $k$ by an eccentricity singular factor, 
$1/(1-e_t^2)^{3k+7/2}$.  Once this term is factored out of the $T_k(e_t)$, the remaining dependence is a polynomial 
in even powers of $e_t$ of order $4(k+1)$, giving each $T_k$ a closed-form expression.  

\begin{widetext}
In what follows, we show that each $T_k(e_t)$ is indeed an energy flux enhancement function that is proportional 
to the (leading log) energy flux at PN order $(3k)L(k)$; i.e., $\mathcal{R}_{(3k)L(k)}(e_t) \propto T_k(e_t)$ 
(further discussion is found in Sec.~\ref{sec:MoreGeneral}).  Therefore, for example, the next two functions in 
this sequence should give ($k=2$) $\mathcal{R}_{6L2}(e_t) \propto T_2(e_t)$ (i.e., the 6PN $\log^2$ term) and
($k=3$) $\mathcal{R}_{9L3}(e_t) \propto T_3(e_t)$ (i.e., the 9PN $\log^3$ term).  If $T_k(e_t)$ represent 
enhancement functions, it should be the case that they all limit to unity for circular orbits.  Then the constant 
of proportionality between $\mathcal{R}_{(3k)L(k)}(e_t)$ and $T_k(e_t)$ will simply be the circular 
orbit flux for the $k$ (integer) order leading-log term.  

We can easily prove that the $T_k(e_t)$ reduce to unity for $e_t=0$ by considering the expansion of 
$g(n,e_t)$ in $e_t$ \cite{ForsEvanHopp16}
\begin{align}
\label{eqn:gexp}
\begin{autobreak}
\MoveEqLeft
g(n,e_t) = \left(\frac{n}{2}\right)^{2n} e_t^{2n - 4} \bigg(
\frac{1}{\Gamma(n-1)^2} - 
\frac{(n-1)(n^2 + 4n -2)}{2 \, \Gamma(n)^2} e_t^2 + 
\frac{6 n^4 + 45 n^3 + 18 n^2 - 48 n + 8}{48 \, \Gamma(n)^2} e_t^4 +
\cdots \bigg) .
\end{autobreak}
\end{align}
Inspection shows that for $n = 1$ the $e_t^{-2}$ and $e_t^0$ coefficients vanish (since 
$\Gamma(0)^{-1} \rightarrow 0$).  The $n = 2$ harmonic is the only one that contributes at $e_t^0$, and its 
coefficient is clearly unity.  For higher harmonics ($n \ge 3$), the expansion begins at $e_t^2$ or higher.  
Thus, in any sum over harmonics of $g(n,e_t)$ times a power of $n/2$ (i.e., some $T_k$), the result is a function 
that equals unity when $e_t=0$. 

As an example of using this process to determine higher-order PN terms, consider the next leading-log term at 
6PN, $\mathcal{R}_{6L2}(e_t)$.  If we introduce the known circular-orbit factor 
$\mathcal{R}_{6L2}^{\rm circ}=366368/11025$ \cite{Fuji12a}, the procedure above suggests that the 
eccentricity-dependent 6PN leading-log flux will be
\begin{align}
\mathcal{R}_{6L2}(e_t) &=\left(\frac{366368}{11025}\right) T_2(e_t) 
=\left(\frac{366368}{11025}\right)\sum_{n=1}^{\infty} \Big(\frac{n^{4}}{16}\Big)g(n,e_t)  
\\ 
&=\frac{366368}{11025(1-e_t^2)^{19/2}}\bigg(1+\frac{16579}{384} e_t^2 +\frac{459595}{1536} e_t^4
+\frac{847853}{1536} e_t^6 +\frac{3672745}{12288} e_t^8 +\frac{1997845}{49152} e_t^{10}
+\frac{41325}{65536} e_t^{12} \bigg) .
\notag 
\end{align}
This closed form expression was, in fact, found in our previous work fitting extremely high precision BHPT numerical 
flux data from a two-dimensional array of orbits to the PN model \eqref{eqn:energxfluxInf} for the energy flux (see 
\cite{MunnETC19} and \textsc{Mathematica} notebook at \cite{BHPTK18}).  (The BHPT data is fit to a model with the 
parameters $y$ and (Darwin) $e$ but as mentioned in Sec.~\ref{sec:secII} for leading-log terms there is no 
difference between those parameters and $x$ and $e_t$ at lowest order in the mass ratio.)  Interestingly, Forseth et 
al.~\cite{ForsEvanHopp16} actually found the entire $\mathcal{R}_{6L2}(e_t)$ term (in their equation (6.13)) but
did not realize that the series terminated at $e_t^{12}$!

In like fashion we can consider the next leading log at integer power of $x$, 9PN $\log^3$.  The circular-orbit 
flux is $\mathcal{R}_{9L3}^{\rm circ}=-(313611008/3472875)$ \cite{Fuji12a}, suggesting that the full 
eccentricity-dependent term is
\begin{align}
 \mathcal{R}_{9L3}(e_t) &=-\left(\frac{313611008}{3472875}\right) T_3(e_t) 
 =-\left(\frac{313611008}{3472875}\right)\sum_{n=1}^{\infty} 
 \Big(\frac{n^{6}}{64}\Big)g(n,e_t)
\notag 
\\
& =-\frac{313611008}{3472875(1-e_t^2)^{25/2}}\bigg(1+\frac{86207}{768} e_t^2 
+\frac{192133}{96} e_t^4
+\frac{21418885}{2048} e_t^6 +\frac{5050405}{256} e_t^8   
\\ 
& \hspace{15.6em}
+\frac{465472553}{32768} e_t^{10}
+\frac{60415733}{16384} e_t^{12} 
+\frac{71973111}{262144} e_t^{14} +\frac{1341375}{524288} e_t^{16} \bigg).
\notag 
\end{align}
This expression also matches perfectly our more recent BHPT numerical fitting results \cite{MunnETC19,BHPTK18}.  
The analogues in the angular momentum flux, $\mathcal{Z}_{6L2}(e_t)$ and $\mathcal{Z}_{9L3}(e_t)$, found 
analytically from the functions $\tilde{T}_2(e_t)$ \eqref{eqn:tildeT2} and $\tilde{T}_3(e_t)$ upon swapping 
$g(n,e_t)$ for $\tilde{g}(n,e_t)$, are easily calculated and have also been shown to match our BHPT numerical 
results. 

With $\mathcal{R}_{0}(e_t)$, $\mathcal{R}_{3L}(e_t)$, $\mathcal{R}_{6L2}(e_t)$, and $\mathcal{R}_{9L3}(e_t)$ 
all determined analytically by this procedure, there is no reason to believe it does not continue 
\emph{ad infinitum}.  Given the circular-orbit flux found by \cite{Fuji12a}, our procedure indicates that the 
$\mathcal{R}_{12L4}(e_t)$ leading-log term will be
\begin{align}
\mathcal{R}&_{12L4}(e_t) = \left(\frac{67112755712}{364651875}\right)T_4(e_t)  
=\left(\frac{67112755712}{364651875}\right) \sum_{n=1}^{\infty} 
 \Big(\frac{n^{8}}{256}\Big)g(n,e_t)
\notag 
\\ 
&=\frac{67112755712}{364651875(1-e_t^2)^{31/2}}\bigg(1+\frac{1667665}{6144} e_t^2
 +\frac{262261909}{24576} e_t^4 +\frac{381097931}{3072} e_t^6 +\frac{4556442679}{8192} e_t^8
 +\frac{141652841401}{131072} e_t^{10} 
\notag 
\\ 
&+\frac{495810570055}{524288} e_t^{12} +\frac{95441646013}{262144} e_t^{14}
 +\frac{233938838161}{4194304} e_t^{16} +\frac{176821654149}{67108864} e_t^{18}
 +\frac{4419580725}{268435456} e_t^{20} \bigg) .
\end{align}

What about still higher-order leading-log terms?  With an understanding of the role of the $T_k(e_t)$, the key 
remaining issue is to determine the general form for the circular-orbit limit of these fluxes.  As it turns out, 
first-order BHPT has the ability to provide the circular-orbit limit of the entire leading-log series.  For 
Schwarzschild EMRIs, BHPT uses spherical harmonics to decompose field and source terms, with mode numbers $l,m$ 
being related to symmetric tracefree mass and current multipole moments like $I_{ij}$.  For eccentric orbits in 
the frequency domain, perturbation quantities become functions of the triple set of mode numbers $l,m,n'$, where 
$n'$ is the Fourier series index in BHPT that gives harmonics of the radial libration frequency.  The index $n'$ 
contrasts with $n$, the power spectrum index in $g(n,e_t)$ and $\tilde{g}(n,e_t)$.  In BHPT, circular orbits 
correspond to $n'=0$, while for the quadrupole moment the circular orbit flux is determined by $n=2$.  Using 
Johnson-McDaniel's $S_{lm}$ tail factorization \cite{JohnMcDa14}, it is possible to use BHPT to extract the 
circular-orbit limit of the entire leading-logarithm series.  Indeed, we can infer from the discussion in Section 
IV of \cite{JohnMcDaShahWhit15} that this limit is generated entirely by the quadrupole factor $| S_{22} |^2$, 
which can be written as
\begin{gather}
| S_{22} |^2 = \exp\bigg[ 2 \bar{\nu} (\gamma_E + 2\log(2) + \log(y)/2) + 4 \pi y^{3/2} +
\sum_{k=2}^{\infty} \frac{\zeta(k)}{k} \Big( (4 y^{3/2} i - \bar{\nu})^k + (-4 y^{3/2} i - \bar{\nu})^k 
- 2 (-2 \bar{\nu})^k \Big)\bigg] .
\label{eqn:S22}
\end{gather}
\end{widetext}
Here, $\bar{\nu} = \nu - l$, where $\nu$ is the \textit{renormalized angular momentum}, an (in general) 
$lmn'$-dependent quantity from the MST analytic function expansion formalism \cite{ManoSuzuTaka96a,ManoSuzuTaka96b} 
of BHPT (note the notational conflict with the symmetric mass ratio).  The parameter $\bar{\nu}$ has a PN expansion 
in powers of $y^3$ ($=x^3$ for our purposes here).  From \eqref{eqn:S22}, the piece that generates the (circular) 
leading logarithms is
\be
\label{eqn:S22LL}
\exp\Big( -\frac{856}{105} \, y^3 \log(y) + 4 \pi y^{3/2} \Big),
\ee
where $-856/105$ is the coefficient of $y^3$ in the PN expansion for $\bar{\nu}$.  Note that this leading-logarithm 
factor is different from one introduced by Damour and Nagar in \cite{DamoNaga07, DamoNaga08}, as theirs 
related to a waveform phase term that cancels in the fluxes.  Eq.~\eqref{eqn:S22LL} immediately yields the 
circular-orbit portion of $\mathcal{R}_{(3k)L(k)}$ as \cite{JohnMcDa15b}
\be
\label{eqn:evencirc}
\mathcal{R}_{(3k)L(k)}^{\rm circ} = \bigg(-\frac{856}{105}\bigg)^{k}\bigg(\frac{1}{k!}\bigg).
\ee
Note that this result exactly matches an earlier estimate given in \cite{Fuji12b} and 
is consistent with that derived through effective field theory arguments in 
\cite{GoldRoss10} (see as well the discussion in \cite{GoldRossRoth14}).

The entire infinite sequence of integer-order leading logarithms can be found by taking the factors 
\eqref{eqn:evencirc} and combining them with the $T_k(e_t)$ summations to yield 
\be
\label{eqn:AllIntLLs}
\mathcal{R}_{(3k)L(k)}(e_t) = \bigg(-\frac{856}{105}\bigg)^{k}\bigg(\frac{1}{k!}\bigg)\sum_{n=1}^{\infty} 
\Big(\frac{n}{2}\Big)^{2k}g(n,e_t)
\ee
for all $k \ge 0$.  These terms are then transformed into closed-form expressions by factoring out the known 
eccentricity singular dependence $1/(1-e_t^2)^{3k+7/2}$ and resumming. 

All of these results carry over to analogously give $\mathcal{Z}_{(3k)L(k)}(e_t)$, since the circular orbit 
limits are the same, $\mathcal{Z}_{(3k)L(k)}^{\rm circ} = \mathcal{R}_{(3k)L(k)}^{\rm circ}$, and only the 
substitution $g(n,e_t) \rightarrow \tilde{g}(n,e_t)$ is required.  Closed-form expressions emerge once the 
singular factors $1/(1-e_t^2)^{2+3k}$ are pulled out.

\subsection{All leading-log enhancement functions at half-integer powers of $x$ are infinite series with known 
coefficients}

To find the leading-log enhancement functions at half-integer powers of $x$, we turn attention to sums over 
$g(n,e_t)$ with odd powers of $n/2$, as mentioned in the Introduction:
\be
\label{eqn:capThetak}
\Theta_k(e_t) = \sum_{n=1}^{\infty} \Big(\frac{n}{2}\Big)^{2k+1}g(n,e_t) ,
\ee
where $k\ge0$ are integers.  Each $\Theta_k(e_t=0) = 1$, just as with the $T_k(e_t)$.  We see immediately that 
one known enhancement function, the 1.5PN tail $\vp(e_t) = \Theta_0(e_t)$, is the first element in this sequence.

Unlike the previous $T_k(e_t)$, the $\Theta_k(e_t)$ functions have a complicated form when translated back to 
the time domain (see e.g., Eq.~4.5 of \cite{ArunETC08a}), and it is strongly suspected \cite{BlanScha93} that 
none will have a closed-form expression in $e_t$.  Nevertheless, each sum provides an infinite series in $e_t^2$ 
with rational coefficients that can be determined rapidly to any order.  Moreover, we can again remove an 
eccentricity singular factor, $1/(1-e_t^2)^{3k+5}$, from each sum that then makes each resummed series converge 
for all $e_t \le 1$.

The prediction is that the sums \eqref{eqn:capThetak} represent the enhancement functions for all leading-log terms 
at half-integer PN orders, not just at 1.5PN.  Each $\Theta_k(e_t)$ is related to the leading-log flux that is 
1.5PN orders higher in the relative flux than the $T_k(e_t)$ with corresponding $k$.  Thus, this class of functions 
will produce the PN terms $\mathcal{R}_{3/2}$, $\mathcal{R}_{9/2L}$, $\mathcal{R}_{15/2L2}$, etc, with each 
constituting the first appearance of a new power of $\log(x)$ at half-integer powers of $x$.  For each $k$ we will 
have $\mathcal{R}_{(3k+3/2)L(k)} \propto \Theta_k$, with the constant of proportionality being again the 
circular-orbit flux.

\begin{widetext}
We consider the specific example of $k=1$ that purports to give $\mathcal{R}_{9/2L}$.  In this case the 
circular-orbit limit is $\mathcal{R}_{9/2L}^{\rm circ} = -3424\pi/105$, which yields
\begin{align}
\mathcal{R}_{9/2L}(e_t) &=-\frac{3424\pi}{105} \sum_{n=1}^{\infty} \Big(\frac{n^{3}}{8}\Big)g(n,e_t)
\\
\notag 
&=-\frac{3424 \pi}{105 (1-e_t^2)^{8}}\bigg(1+\frac{19555}{768} e_t^2 
+\frac{303647}{3072} e_t^4
+\frac{13263935}{147456} e_t^6 
+\frac{64393025}{3538944} e_t^8
+\frac{557011627}{1415577600} e_t^{10} 
+\cdots\bigg) .
\end{align}
The expansion for $\mathcal{R}_{9/2L}$ matches perfectly the results from fitting, to $e_t^{18}$ as found in 
\cite{ForsEvanHopp16} and to $e_t^{30}$ as obtained in our more recent work \cite{MunnETC19,BHPTK18}.
The non-singular infinite series converges to approximately $233.8451300137$ as $e_t \rightarrow 1$.
In the same way, $\Theta_2(e_t)$ can be evaluated to reproduce $\mathcal{R}_{15/2L2}(e_t)$, which we found 
matches our BHPT fitting results to $e_t^{30}$ \cite{MunnETC19,BHPTK18}.

Rather than enumerate explicitly added individual leading-log functions, we jump straight to the form of the 
general solution.  Once again, \eqref{eqn:S22LL} provides the circular-orbit limit to the leading-log energy 
fluxes, which for the half-integer power in $x$ sequence is
\be
\label{eqn:oddcirc}
\mathcal{R}_{(3k+3/2)L(k)}^{\rm circ} = 4\pi\bigg(-\frac{856}{105}\bigg)^{k}\bigg(\frac{1}{k!}\bigg) .
\ee
The only difference from the previous sequence being the added factor of $4\pi$.  The circular-orbit limits
can be combined with \eqref{eqn:capThetak} to yield the full set ($k \ge 0$) of half-integer in $x$ leading-log 
energy fluxes
\be
\label{eqn:AllHalfIntegerLLs}
\mathcal{R}_{(3k+3/2)L(k)}(e_t) = \bigg(\frac{4\pi}{k!}\bigg)\bigg(-\frac{856}{105}\bigg)^{k}\sum_{n=1}^{\infty} 
\Big(\frac{n}{2}\Big)^{2k+1}g(n,e_t) .
\ee
Each term will have a singular behavior like $1/(1-e_t^2)^{3k+5}$ as $e_t\rightarrow 1$.  Once these factors 
are pulled out, each resummed series will converge as $e_t\rightarrow 1$, though none of them is expected to 
truncate and leave a polynomial.  The series coefficients are known in the sense that they can easily be calculated 
analytically from \eqref{eqn:capThetak} and \eqref{eqn:gfunc} with minimal symbolic computational expense. 

The results carry over from \eqref{eqn:AllHalfIntegerLLs} to give the corresponding leading-log angular momemtum 
fluxes $\mathcal{Z}_{(3k+3/2)L(k)}(e_t)$ by doing nothing more than substituting $\tilde{g}(n,e_t)$ in place of 
$g(n,e_t)$.  The eccentricity singular factors in this case will be $1/(1-e_t^2)^{3k+7/2}$.

\subsection{Summary}

We have shown that the eccentricity dependence of the entire infinite sequence of leading-logarithm energy and 
angular momentum PN flux terms is analytically determined by the Newtonian quadrupole moment spectra $g(n,e_t)$ 
and $\tilde{g}(n,e_t)$.  This implies further that all of the leading-log terms appear only at lowest order in the 
mass ratio $\nu$.  In the next section we show that additional analytic knowledge of terms at high PN order, this 
time of the eccentricity dependence of the subleading logarithms, can be coaxed out of a combination of information 
in the Newtonian quadrupole moment power spectra and BHPT flux results.

\section{Additional PN structure from $g(n,e_t)$ and perturbation theory}
\label{sec:additional6L}

\subsection{Generalizations of $\chi(e_t)$}
\label{sec:generalChi}

As the previous section argued, the succession of Newtonian mass quadrupole sums \eqref{eqn:capTk} and 
\eqref{eqn:capThetak} provides the eccentricity dependence of the entire leading-log PN sequence.  The first three 
elements in this sequence were equal to, or proportional to, the previously known flux functions 
$\mathcal{R}_{0}(e_t)$, $\mathcal{R}_{3/2}(e_t)$, and $\mathcal{R}_{3L}(e_t)$.  There was, however, one other 
previously known enhancement function, $\chi(e_t)$, that did not make an appearance within the leading-log 
sequence.  Instead, as inspection of \eqref{eqn:3PNE} indicates, $\chi(e_t)$ showed up as part of 
$\mathcal{R}_{3}(e_t)$, the non-log part at 3PN order, which we classify as a subleading log.  As the Introduction 
outlined, this hints at the possible use of two more classes of sums, namely
\be
\label{eqn:lambXi}
\Lambda_k(e_t) =\sum_{n=1}^{\infty} \Big(\frac{n}{2}\Big)^{2k} \log\Big(\frac{n}{2}\Big) \, g(n,e_t),
\qq \qq \Xi_k(e_t) =\sum_{n=1}^{\infty} \Big(\frac{n}{2}\Big)^{2k+1} \log\Big(\frac{n}{2}\Big) \, g(n,e_t),
\ee
for integers $k\ge1$.  It is clear that $\Lambda_1(e_t)$ reproduces the 3PN enhancement function $\chi(e_t)$.

A first question to ask is, if more of these functions were to appear in the PN expansion, at what PN order 
would they show up?  We can answer that question by considering their divergence properties as $e_t\rightarrow 1$.
As stated in Sec.~\ref{sec:secII}, $\chi(e_t)$ contains the logarithmic divergence found in 
$-(3/2) F(e_t) \log(1-e_t^2)$ in addition to the algebraic singularity of $F(e_t)$.  A similar behavior appears 
in each $\Lambda_k(e_t)$ and $\Xi(e_t)$.  To see this, we apply the same asymptotic analysis found in Sec.~IV 
of \cite{ForsEvanHopp16}, using the transition zone asymptotic expansions of $J_n(n e_t)$ (i.e., large $n$ with 
$e_t \simeq 1$ \cite{DLMF}) to expand $g(n,e_t)$ and replacing the sum over $n$ with an integral over a continuous 
variable $\xi = \rho(z) n$.  Here, $\rho(z) = \log\left(\frac{1 + \sqrt{z}}{\sqrt{1-z}}\right)-\sqrt{z}$ and 
$z = 1 - e_t^2$.  Then the log terms in \eqref{eqn:lambXi} are replaced by
\be
\log\left(\frac{n}{2}\right) \rightarrow \log\left(\frac{\xi}{2 \rho(z)}\right) ,
\ee
followed by splitting off the $-\log(\rho)$ portion, expanding in $z$, and integrating over $\xi$.  The result is 
that we find the asymptotic singular dependence of $\Lambda_k(e_t)$ and $\Xi(e_t)$ to be 
\begin{align}
\label{eqn:asympLambda}
\Lambda_k(e_t) &\sim -\frac{3}{2} T_k(e_t) \log(1-e_t^2) 
\sim \Lambda_k^{(0)} \, \log(1-e_t^2) \, (1-e_t^2)^{-3k-7/2} , 
\\
\label{eqn:asympXi}
\Xi(e_t) &\sim -\frac{3}{2} \Theta_k(e_t) \log(1-e_t^2) 
\sim \Xi_k^{(0)} \, \log(1-e_t^2) \, (1-e_t^2)^{-3k-5} ,
\end{align}
respectively, where $\Lambda_k^{(0)}$ and $\Xi_k^{(0)}$ are constants.  The algebraic part of the eccentricity 
singular dependence indicates that, if these terms show up in the PN fluxes at all, they will appear at relative 
PN orders $3k$ and $3k + 3/2$, respectively.\footnote{The same conclusion can easily be reached by power counting, 
since each power of $n$ in \eqref{eqn:lambXi} corresponds to a factor of $\Omega_r$ from time derivatives of 
$I_{ij}$.  Thus each power of $n$ brings with it a factor proportional to $x^{3/2}$, at lowest order in $\nu$, 
making the relative PN orders $3k$ and $3k + 3/2$ as mentioned.  The asymptotic analysis, however, has the advantage 
of also revealing the logarithmic singularity and (importantly) the connections to the previously defined functions 
$T_k(e_t)$ and $\Theta_k(e_t)$.}  Given that these functions do not show up in the leading-log 
sequence, but based on the way $\chi(e_t)$ appears in $\mathcal{R}_{3}$, a conjecture would be that they 
contribute to the subleading-log sequence (previously defined).  Thus, with the reemergence of $T_k(e_t)$ 
in \eqref{eqn:asympLambda}, we might expect $\Lambda_k(e_t)$ to contribute to the subleading-log sequence 
$\mathcal{R}_{3}$, $\mathcal{R}_{6L}$, $\mathcal{R}_{9L2}$, etc.  Likewise, since $\Theta_k(e_t)$ reappears in
\eqref{eqn:asympXi}, we conjecture that the $\Xi_k(e_t)$ contribute to the half-integer subleading-log 
sequence $\mathcal{R}_{9/2}$, $\mathcal{R}_{15/2L}$, $\mathcal{R}_{21/2L2}$, etc.  Furthermore, the 
asymptotic connection between $\Lambda_k(e_t)$ and $T_k(e_t)$ in \eqref{eqn:asympLambda} leads us to 
conjecture that the higher order subleading-log terms $\mathcal{R}_{(3k)L(k-1)}(e_t)$ all have structures nearly 
identical to that of $\mathcal{R}_{3}(e_t)$ \eqref{eqn:3PNE}, with closed-form expressions supplementing the 
appearance of $\Lambda_k(e_t)$.

We note in passing that there is another way of regarding subleading-log terms.  These terms, which appear at 
PN order $3k$ or $3k+3/2$ but involve one power of $\log(x)$ less than the leading-log term, can also be thought 
of as 3PN \emph{corrections} to the previous leading-log in the series.  Thus, $\mathcal{R}_{3}(e_t)$, 
$\mathcal{R}_{9/2}(e_t)$, $\mathcal{R}_{6L}(e_t)$, and $\mathcal{R}_{15/2L}(e_t)$ are 3PN corrections to 
$\mathcal{R}_{0}(e_t)$, $\mathcal{R}_{3/2}(e_t)$, $\mathcal{R}_{3L}(e_t)$, and $\mathcal{R}_{9/2L}(e_t)$, 
respectively.  This alternative designation scheme will become especially useful in future work, as we compute
additional sequences of logarithms in the two flux expansions.

\subsection{The 6PN subleading-log example}

The conjectures made in the previous subsection appear to be correct, as far as we have been able to verify with 
BHPT calculations.  To give an example and demonstrate the structure of a subleading-log term beyond 
$\mathcal{R}_{3}(e_t)$, we consider $\mathcal{R}_{6L}(e_t)$.  In the end, we obtain the entire 6L term (i.e., 
its entire $e_t$ dependence) at lowest order in $\nu$.  Because our analysis makes heavy use of BHPT results, we 
work initially in terms of Darwin eccentricity $e$ and compactness $y$.  We first express $\Lambda_2(e_t)$ and 
$T_2(e_t)$ in terms of $e$, as these functions are needed in the analysis.  However, since they only depend upon 
the Newtonian mass quadrupole spectrum, they can be converted by simply swapping $e_t$ for $e$. 

The process then involves (i) making an ansatz on the analytic form of $\mathcal{L}_{6L}(e)$ that includes an 
assumed dependence on $\Lambda_2(e)$ and $T_2(e)$, 
(ii) using BHPT to compute analytic coefficients in the expansion of $\mathcal{L}_{6L}(e)$ to a high finite order in $e^2$ (in our case, this was done using high-precision numerical data and ``experimental 
mathematics''; see \cite{JohnMcDaShahWhit15,ForsEvanHopp16,MunnETC19}) for details), 
(iii) subtracting the parts involving $\Lambda_2(e)$ 
and $T_2(e)$ to determine the (closed-form algebraic) rest of the analytic model, and (iv) converting back 
to $e_t$ to obtain $\mathcal{R}_{6L}(e_t)$.  

The guess for the general form of $\mathcal{L}_{6L}(e)$, based on resemblance to \eqref{eqn:3PNE}, is
\begin{align}
\label{eqn:3PNform}
&\mathcal{L}_{6L}^\text{model} = 
\frac{1}{(1-e^2)^{19/2}}\biggl[
a_0+a_2e^2+a_4 e^4+a_6 e^6+a_8 e^8+a_{10} e^{10}+a_{12} e^{12}+a_{14} e^{14} + \sqrt{1-e^2}\bigg(b_0+b_2e^2+b_4 e^4 
\notag 
\\ 
&\qq +b_6 e^6  +b_8 e^8+b_{10} e^{10}+ b_{12} e^{12}\bigg) \biggr]
+\left[c_1\pi^2+c_2\gamma_\text{E}+c_3\log(2)
+c_4\log\left(\frac{1-e^2}{1+\sqrt{1-e^2}}\right)\right]T_{2}(e)
+d_1\Lambda_{2}(e),
\end{align}
for some rational coefficient set $\{a_i, b_i, c_i, d_i\}$.  In the model, $T_2$ reappears but is written as a 
function of $e$
\begin{align}
T_2(e) &=
\frac{1}{(1-e^2)^{19/2}}\bigg(1+\frac{16579}{384} e^2+\frac{459595}{1536} e^4+\frac{847853}{1536} e^6
+\frac{3672745}{12288}e^8 +\frac{1997845}{49152}e^{10} + \frac{41325}{65536} e^{12}\bigg) ,
\end{align}
and so does $\Lambda_2$, also written in terms of $e$
\begin{align}
\Lambda_2(e)&=\frac{1}{(1-e^2)^{19/2}}\bigg[ 
\left(-\frac{22147 \log (2)}{384} +\frac{59049 \log (3)}{1024}\right)e^2+
 \left(\frac{945063 \log (2)}{512}-\frac{3365793 \log   (3)}{4096}\right)e^4  \notag \\
&+ \left(-\frac{47071565 \log (2)}{1536}+\frac{357108669 \log (3)}{65536}
+\frac{6103515625 \log (5)}{589824}\right)e^6 \notag \\
&+\left(\frac{10209340261 \log (2)}{36864}+\frac{27480125205 \log (3)}{524288}
-\frac{726318359375 \log (5)}{4718592}\right)e^8 +\cdots\bigg] .
\end{align}
For brevity only the first part of $\Lambda_2(e)$ is presented, despite having been (necessarily) determined to 
$e^{30}$.  Also, it is not necessary to isolate the logarithmic divergence in $\Lambda_2(e)$.  Despite the generality 
of \eqref{eqn:3PNform}, we anticipate some coefficients being linked.  Based on the form of $\mathcal{R}_3$ and 
the structure found within the 220 mode flux (see \cite{MunnETC19} and Sec.~\ref{sec:MoreGeneral}), we expected 
(and ultimately confirmed) the following connections: $c_2 = c_3/3 = c_4 = d_1$.

The next step is computation of the analytic expansion of $\mathcal{L}_{6L}(e)$ through $e^{30}$, which
was done using high-precision BHPT numerical data, fitting \cite{MunnETC19} to the PN model, and using the PSLQ 
integer relation algorithm \cite{FergBailArno99}.  That process yielded 
\begin{align}
\begin{autobreak}
\MoveEqLeft
 \mathcal{L}_{6L}^{(30)} = \frac{1}{(1-e^2)^{19/2}}\bigg[-\frac{246137536815857}{314659144800}
+\frac{1465472\gamma_E}{11025}-\frac{13696\pi^2}{315}
+\frac{2930944\log (2)}{11025}
+ \bigg(-\frac{25915820507512391}{629318289600}
+\frac{189812971\gamma_E}{33075}
-\frac{1773953\pi^2}{945}
+\frac{18009277\log (2)}{4725}
+\frac{75116889\log (3)}{9800}\bigg)e^{2}
+ \bigg(-\frac{56861331626354501}{167818210560}
+\frac{1052380631\gamma_E}{26460}-\frac{9835333\pi^2}{756}
+\frac{42983885171\log (2)}{132300}-\frac{4281662673\log (3)}{39200}\bigg)e^{4}
+ \bigg(-\frac{710806279550045831}{1006909263360}
+\frac{9707068997\gamma_E}{132300}
-\frac{90720271\pi^2}{3780}-\frac{519508209691\log (2)}{132300}
+\frac{454281905709\log (3)}{627200}+\frac{2795166015625\log (5)}{2032128}\bigg)e^{6}
+ \bigg(-\frac{10213351238593603069}{40276370534400}
+\frac{8409851501\gamma_E}{211680}-\frac{78596743\pi^2}{6048}
+\frac{117139032193219\log (2)}{3175200}
+\frac{6991554521601\log (3)}{1003520}-\frac{47517822265625\log (5)}{2322432}\bigg)e^{8}
+ \bigg(\frac{3985515397336843519}{26850913689600}
+\frac{4574665481\gamma_E}{846720}
-\frac{42753883\pi^2}{24192}-\frac{252510878807655859\log (2)}{952560000}
-\frac{576360297584196039\log (3)}{4014080000}
+\frac{223101765869140625\log (5)}{1560674304}
+\frac{380483822091361849\log (7)}{6635520000}\bigg)e^{10}
+ \bigg(\frac{50719954422267749}{3254656204800}
+\frac{6308399\gamma_E}{75264}-\frac{294785\pi^2}{10752}
+\frac{2887481794238961637\log (2)}{1270080000}
+\frac{17322463230547056201\log (3)}{16056320000}
-\frac{1297619485595703125\log (5)}{2080899072}
-\frac{2663386754639532943\log (7)}{2949120000}\bigg)e^{12}
+ \bigg(-\frac{477961162088755717}{14320487301120}
-\frac{339392544622900323521\log (2)}{17503290000}
-\frac{15568492847979888930357\log (3)}{6294077440000}
+\frac{20971917520162841796875\log (5)}{11012117889024}
+\frac{77148041218710802588787\log (7)}{11466178560000}\bigg)e^{14}
+\cdots+\kappa_{30}e^{30} \bigg] .
\end{autobreak}
\end{align}
The truncated expansion is distinguished by the superscript $(30)$.  Once again an abbreviation of the full series 
is presented; the placeholder coefficient $\kappa_{30}$ denotes the true length of the analytic expansion.  The 
full series to $e^{30}$ would require multiple pages to print out.  

We continue the procedure by subtracting off the piece in the ansatz with no closed-form expression, 
namely $\Lambda_2(e)$.  The proportionality constant is $d_1 = 1465472/11025$, easily found through inspection 
of the $\mathcal{L}_{6L}^{(30)}$ series.  Once $\Lambda_2(e)$ is removed, a significant reduction in complexity 
is observed, which allows the \emph{entire} remaining series to be written down through $e^{30}$
\begin{align}
\begin{autobreak}
\MoveEqLeft
\mathcal{L}_{6L}^{(30)}-\frac{1465472}{11025}\Lambda_2(e) = 
\frac{1}{(1-e^2)^{19/2}}\bigg[-\frac{246137536815857}{314659144800}
+\frac{1465472 \gamma_E }{11025}-\frac{13696 \pi ^2}{315}+\frac{2930944 \log (2)}{11025}
+\left(-\frac{25915820507512391}{629318289600}+\frac{189812971 \gamma_E }{33075} -\frac{1773953 \pi^2}{945}+\frac{379625942 \log(2)}{33075}\right) e^2 
+ \left(-\frac{56861331626354501}{167818210560}+\frac{1052380631 \gamma_E }{26460} -\frac{9835333 \pi^2}{756}+\frac{1052380631 \log (2)}{13230}\right) e^4
+ \left(-\frac{710806279550045831}{1006909263360}+\frac{9707068997 \gamma_E}{132300} -\frac{90720271 \pi^2}{3780}+\frac{9707068997 \log (2)}{66150}\right) e^6
+\left(-\frac{10213351238593603069}{40276370534400}+\frac{8409851501 \gamma_E }{211680} -\frac{78596743 \pi^2}{6048}+\frac{8409851501 \log(2)}{105840}\right) e^8
+ \left(\frac{3985515397336843519}{26850913689600}+\frac{4574665481 \gamma_E }{846720} -\frac{42753883 \pi^2}{24192}+\frac{4574665481 \log(2)}{423360}\right) e^{10}
+\left(\frac{50719954422267749}{3254656204800}+\frac{6308399 \gamma_E}{75264} -\frac{294785 \pi^2}{10752}+\frac{6308399 \log (2)}{37632}\right) e^{12}
-\frac{477961162088755717 }{14320487301120}e^{14}-\frac{5413490909883323 }{182078668800}e^{16}
-\frac{5584575351395413 }{218494402560}e^{18}-\frac{81136058237959211}{3641573376000}e^{20}
-\frac{1578479509403151527 }{80114614272000}e^{22}
-\frac{2261257978156608611 }{128183382835200}e^{24}
-\frac{531918812054997639011 }{33327679537152000}e^{26}
-\frac{388387963969333233793 }{26662143629721600}e^{28}
-\frac{892815371640935597927 }{66655359074304000}e^{30} \bigg] .
\end{autobreak}
\label{eqn:6L30minusLambda}
\end{align}
We note also that each coefficient after $e^{12}$ is purely rational.  The undeniable conclusion is that 
$\Lambda_2(e)$ does indeed provide a desired contribution to $\mathcal{L}_{6L}(e)$. 

In the next step, we confirm another tenent of the analytic model---that all of the transcendental numbers, 
$\gamma_E, \pi^2,$ and $\log(2)$, in the first terms up to $e^{12}$ in \eqref{eqn:6L30minusLambda} simply appear 
as a specific combination that multiplies $T_2(e)$ (a function which contains a 12th order polynomial).  The revised 
model then becomes
\begin{align}
\mathcal{L}_{6L}^\text{model} &= 
\frac{1}{(1-e^2)^{19/2}}\biggl[
a_0+a_2e^2+a_4 e^4+a_6 e^6+a_8 e^8+a_{10} e^{10}+a_{12} e^{12} + a_{14} e^{14}
\notag
\\
& \qq \qq
+\sqrt{1-e^2}\bigg(b_0+b_2e^2+b_4 e^4+b_6 e^6+b_8 e^8
+b_{10} e^{10} +b_{12} e^{12}\bigg) 
\biggr]
\notag
\\
& +\left[\frac{1465472}{11025}\gamma_\text{E}-
\frac{13696 \pi ^2}{315}+\frac{4396416}{11025}\log(2)
+\frac{1465472}{11025}\log\left(\frac{1-e^2}{1+\sqrt{1-e^2}}\right) \right]T_2(e) 
+\frac{1465472}{11025}\Lambda_2(e) ,
\end{align}
once the $c_i$ coefficients are determined and inserted.  If we now subtract the $T_2(e)$ part of the model as 
well from $\mathcal{L}_{6L}^{(30)}$ (i.e., from \eqref{eqn:6L30minusLambda}), we are left with
\begin{align}
\label{eqn:E6LRat}
\begin{autobreak}
\MoveEqLeft
\frac{1}{(1-e^2)^{19/2}}\bigg(-\frac{246137536815857}{314659144800}-\frac{5170616505141979 }{125863657920}e^{2}-\frac{280649774449416601 }{839091052800}e^{4}
-\frac{3391928161684113811 }{5034546316800}e^{6}
-\frac{1456012194152323001 }{8055274106880}e^{8}
+\frac{29600878702417369091 }{134254568448000}e^{10}
+\frac{1074387193648790113 }{16273281024000}e^{12}
+\frac{17814341408826553  }{4773495767040}e^{14}
-\frac{31846235946197 }{303464448000}e^{16}
-\frac{219944663655131 }{273118003200}e^{18}
-\frac{113553895395893 }{115605504000}e^{20}
-\frac{172257218309077 }{173408256000}e^{22}
-
\frac{394386143943349 }{416179814400}e^{24}
-\frac{700775531336071  }{792723456000}e^{26}
-\frac{25403642219761117 }{31074759475200}e^{28}
-\frac{19524067936619881 }{25895632896000}e^{30} \bigg) ,
\end{autobreak}
\end{align}
a purely rational series in $e^2$.

At this point the 16 rational coefficients in \eqref{eqn:E6LRat} must be determined, if possible, by the remaining 
15 unknown constants $a_i$ and $b_i$ in the model.  This was the reason for carrying out our numerical fitting 
and analytic expansions to $e^{30}$, to provide an overdetermined system of equations.  We find that indeed 
a solution for the $a_i$ and $b_i$ can be obtained, verifying the ansatz and giving the entire analytic 
structure of $\mathcal{L}_{6L}(e)$ as
\begin{align}
\mathcal{L}_{6L}(e) &= 
\frac{1}{(1-e^2)^{19/2}}
\biggl[-\frac{2634350510203129}{1573295724000}-\frac{239953038071655043 }{3146591448000}e^{2}
-\frac{411009526770805477 }{839091052800}e^{4} 
\notag 
\\
&-\frac{17212115479135988207 }{25172731584000}e^{6} 
-\frac{81213393300931861 }{40276370534400}e^{8}
+\frac{6299935941231102319 }{26850913689600}e^{10}
+\frac{30953812320468361 }{650931240960}e^{12}  
\notag 
\\
&+\frac{205680487293493 }{227309322240}e^{14}
+\sqrt{1-e^2} \bigg(\frac{74362302719}{83349000}+\frac{5938296687287 }{166698000}e^{2}
+\frac{1203568974373 }{6945750}e^{4}
\notag 
\\
&+\frac{67465356696233 }{666792000}e^{6} 
-\frac{1111945369132247 }{10668672000}e^{8} 
-\frac{32687662125259 }{790272000}e^{10}-\frac{116022069 }{100352}e^{12}\bigg)
\biggr] 
\notag 
\\
&+ \left[\frac{1465472}{11025}\gamma_\text{E}-
\frac{13696 \pi ^2}{315}+\frac{4396416}{11025}\log(2)
+\frac{1465472}{11025}\log\left(\frac{1-e^2}{1+\sqrt{1-e^2}}\right)
\right]T_2(e) +\frac{1465472}{11025}\Lambda_2(e) .
\label{eqn:sixLogExact}
\end{align}
Everything in this expression for $\mathcal{L}_{6L}(e)$ is in closed form except for the infinite series 
$\Lambda_2(e)$, which nevertheless itself has coefficients that can be easily determined analytically to arbitrary 
order in $e^2$.  

Having achieved this end in the energy flux, we can perform precisely the same procedure on the 6L angular 
momentum flux term to find
\begin{align}
\mathcal{J}_{6L}&(e) = 
\frac{1}{(1-e^2)^8}\biggl[-\frac{2460815702382469}{1573295724000}-\frac{60681012190195757 }{1573295724000}e^2
-\frac{613664666042477719}{4195455264000}e^4 
\notag
\\
&-\frac{142507823837043079 }{1258636579200}e^6 
+\frac{220635683492763683 }{40276370534400}e^8
+\frac{1157237897488423}{114747494400}e^{10}
+\frac{39115865356031}{113654661120}e^{12} 
\notag
\\
&+\sqrt{1-e^2} \bigg(\frac{86202239}{110250}+\frac{2193242627}{147000}e^2
 +\frac{31184553527}{882000}e^4-\frac{20643131927}{3528000}e^6-\frac{190378390633 }{14112000}e^8
-\frac{8199949}{12544}e^{10}\bigg) \biggr] 
\notag
\\
&+\left[\frac{1465472}{11025}\gamma_\text{E}-
\frac{13696 \pi ^2}{315}+\frac{4396416}{11025}\log(2)
+\frac{1465472}{11025}\log\left(\frac{1-e^2}{1+\sqrt{1-e^2}}\right)
\right]\tilde{T}_2(e) 
+\frac{1465472}{11025}\tilde{\Lambda}_{2}(e) , 
\end{align}
where the (closed-form) enhancement function 
\be
\label{eqn:tildeT2}
\tilde{T}_2(e) = \frac{1}{(1-e^2)^8} 
\bigg(1+\frac{3259 }{128}e^{2}+\frac{1581 }{16}e^{4}+\frac{46015 }{512}e^{6}+\frac{18595 }{1024}e^{8}
+\frac{6345 }{16384}e^{10}\bigg) , \qq \qq \qq \qq
\ee
is used and where the leading part of the infinite series for $\tilde{\Lambda}_2(e)$ is
\begin{align}
\tilde{\Lambda}_2(e) &=
\frac{1}{(1-e^2)^8} \bigg[\left(-\frac{4923 \log (2)}{128} +\frac{19683 \log (3)}{512}\right) e^2
+\left(\frac{16037 \log (2)}{16}-\frac{1003833 \log (3)}{2048}\right) e^4  
\notag 
\\
&+\left(-\frac{63030583 \log (2)}{4608}+\frac{94458717 \log (3)}{32768}
+\frac{1220703125 \log (5)}{294912}\right) e^6 
\notag 
\\
&+\left(\frac{976014461 \log (2)}{9216}+\frac{3811868829 \log (3)}{262144}
-\frac{130615234375 \log (5)}{2359296}\right) e^8 + \cdots \bigg] ,
\end{align}
though for our purposes (again) it had to be expanded to $e^{30}$.  Note that the $c_i$ and $d_1$ coefficients are
exactly the same as those in the 6L energy flux.

With complete understanding of $\mathcal{L}_{6L}(e)$ and $\mathcal{J}_{6L}(e)$ (in terms of PN parameters 
$e$ and $y$), we can obtain $\mathcal{R}_{6L}(e_t)$ and $\mathcal{Z}_{6L}(e_t)$ (at lowest order in $\nu$) by 
using $y = x + \mathcal{O}(\nu)$ and converting $e$ to $e_t$ using \cite{ForsEvanHopp16}
\be
\label{eqn:eToet}
\frac{e^2}{e_t^2} = 1+6 y+\frac{17-21 e_t^2+15 \sqrt{1-e_t^2}}{1-e_t^2} y^2 +\frac{26-107 e_t^2 + 54 e_t^4
+\left(150 - 90 e_t^2 \right)\sqrt{1-e_t^2}}{\left(1-e_t^2\right)^2} y^3 + \mathcal{O}(y^4) .
\ee
The effect of this PN expansion between $e$ and $e_t$ is that, in order to convert to $\mathcal{R}_{6L}(e_t)$ from 
$\mathcal{L}_{6L}(e)$, we have to account for terms that ripple through from also transforming $\mathcal{L}_{3L}(e)$, 
$\mathcal{L}_{4L}(e)$, and $\mathcal{L}_{5L}(e)$.  To accomplish this, each of these flux terms must be known to 
$e^{30}$ (see \cite{MunnETC19,BHPTK18}).  The same procedure is followed to convert to $\mathcal{Z}_{6L}(e_t)$ from 
$\mathcal{J}_{6L}(e)$.  We find
\begin{align}
\label{eqn:R6L}
\mathcal{R}_{6L}(e_t) &= 
\frac{1}{(1-e_t^2)^{19/2}}\biggl[-\frac{2634350510203129}{1573295724000}
-\frac{76144416345305443 }{3146591448000}e_t^{2}-\frac{31937513191666597 }{839091052800}e_t^{4}  \notag \\
&-\frac{399990451980530207 }{25172731584000}e_t^{6}-\frac{2328285213193351381 }{40276370534400}e_t^{8}
-\frac{821024946321249521 }{26850913689600}e_t^{10}
-\frac{113510030676997 }{59175567360}e_t^{12} \notag \\
&+\frac{732785694853 }{227309322240}e_t^{14}
+\sqrt{1-e_t^2} \bigg(\frac{74362302719}{83349000}-\frac{1295489312713 }{166698000}e_t^{2}-\frac{9312957259141 }{55566000}e_t^{4}-\frac{220905190597267 }{666792000}e_t^{6} \notag \\
&-\frac{1481390282809247 }{10668672000}e_t^{8}-\frac{8130086922259 }{790272000}e_t^{10}
-\frac{10593 }{448}e_t^{12} \bigg) \biggr] 
\\
\notag 
&+\left[\frac{1465472}{11025}\gamma_\text{E}-
\frac{13696 \pi ^2}{315}+\frac{4396416}{11025}\log(2)
+\frac{1465472}{11025}\log\left(\frac{1-e_t^2}{1+\sqrt{1-e_t^2}}\right)
\right]T_2(e_t) 
+\frac{1465472}{11025}\Lambda_2(e_t),
\end{align}
\begin{align}
\mathcal{Z}_{6L}(e_t) &= 
\frac{1}{(1-e_t^2)^8}\biggl[-\frac{2460815702382469}{1573295724000}-\frac{14809210436217557}{1573295724000}e_t^{2}+\frac{38156471442639881 }{4195455264000}e_t^{4}
+\frac{489605424663941 }{1258636579200}e_t^{6} \notag \\
&-\frac{530424582265919197 }{40276370534400}e_t^{8}-\frac{153117422046377 }{114747494400}e_t^{10}+
\frac{121354621781 }{37884887040}e_t^{12}
+\sqrt{1-e_t^2} \bigg(\frac{86202239}{110250}-\frac{1047437123 }{147000}e_t^{2}
\notag 
\\
& -\frac{54935631223 }{882000}e_t^{4}-\frac{189779591177 }{3528000}e_t^{6}
 -\frac{93801917383 }{14112000}e_t^{8}-\frac{2461 }{112}e_t^{10}\bigg) \biggr] 
\\
\notag 
&+\left[\frac{1465472}{11025}\gamma_\text{E}-
\frac{13696 \pi ^2}{315}+\frac{4396416}{11025}\log(2)
+\frac{1465472}{11025}\log\left(\frac{1-e_t^2}{1+\sqrt{1-e_t^2}}\right)
\right]\tilde{T}_2(e_t) 
+\frac{1465472}{11025}\tilde{\Lambda}_{2}(e_t).
\end{align}
In principle this procedure might be followed to simplify and make analytically known the next 
subleading-log terms (at an integer power of $x$), i.e., $\mathcal{L}_{9L2}$ and $\mathcal{J}_{9L2}$.

\end{widetext}
\subsection{The 9/2PN subleading-log example}

The procedure laid out above for using the Newtonian quadrupole to determine the subleading-log term 
$\mathcal{L}_{6L}(e)$, at an integer power of $y$, also works at half-integer powers of $y$.  The first 
such term would be the subleading-log $\mathcal{L}_{9/2}$ (associated with leading-log $\mathcal{L}_{9/2L}$).   
Recall that we can also consider this term to be a 3PN correction to the previous leading-log, 
$\mathcal{L}_{3/2}(e)$.  Since the 1.5PN tail $\mathcal{L}_{3/2}(e)$ is an infinite series, we must expect 
$\mathcal{L}_{9/2}$ to be one as well.  We show here, however, that if we follow the same procedure and isolate 
the transcendental portion (except for an overall multiplicative factor of $\pi$) using the Newtonian mass quadrupole 
sums $\Theta_{1}(e)$ and $\Xi_1(e)$, then the remaining infinite series involves only rational coefficients.  We 
thus transform the complicated fitting result in \cite{MunnETC19,BHPTK18} into a much more manageable form
\begin{widetext}
\begin{align}
\mathcal{L}_{9/2}(e) &= 
\frac{\pi}{(1-e^2)^{8}}\biggl[\frac{265978667519}{745113600}+\frac{5009791040801}{447068160} e^2
+\frac{4046503446057439}{71530905600} e^4 +\frac{551321612915453}{8047226880} e^6 
\notag 
\\
&+\frac{422210831769796213}{65922882600960} e^8
-\frac{18560339255510812003}{2746786775040000} e^{10}
-\frac{146292481172437451857}{339031967662080000} e^{12}    
\notag 
\\
& +\frac{392821388634552281893}{5285816586731520000} e^{14}
+\frac{2162084778435646377506023}{17011268009412526080000} e^{16}
+\frac{140095355726033870461460573}{1071709884592989143040000} e^{18} 
\notag 
\\
& +\frac{943121499884145402173125024543}{7716311169069521829888000000} e^{20} 
+\frac{741566762964436290955111519639}{6669097510410086724403200000} e^{22} 
\notag 
\\
& +\frac{863925808693107071875922125163041313}{8604736371831510295293984768000000} e^{24}
+\frac{26361076468942343108164030017209652079}{290840089367905047980936685158400000} e^{26}
+ \cdots  \biggr] 
\notag 
\\
& - \left[\frac{6848\pi}{105} \gamma_\text{E} + \frac{20544\pi}{105} \log(2)
+ \frac{6848\pi}{105} \log\left(\frac{1-e^2}{1+\sqrt{1-e^2}}\right)
\right]\Theta_1(e) - \frac{6848 \pi}{105}\Xi_1(e) .
\label{eqn:4p5Compact}
\end{align}
\end{widetext}
While \eqref{eqn:4p5Compact} is still an infinite series, we have identified some of the tail dependence 
by isolating the entire transcendental portion of $\mathcal{L}_{9/2}$ using only the Newtonian mass quadrupole. 
The process translates trivially from energy to angular momentum fluxes.  Furthermore, the route followed in 
the previous subsection could be used again to translate $\mathcal{L}_{9/2}(e)$ to $\mathcal{R}_{9/2}(e_t)$. 
Finally, with enough BHPT fitting data, similar simplifications could be performed at higher PN orders, for 
$\mathcal{L}_{15/2L}$, $\mathcal{J}_{15/2L}$, $\mathcal{L}_{21/2L2}$, $\mathcal{J}_{21/2L2}$, etc. 

\subsection{Discussion}

Separating off the transcendentals, as done in \eqref{eqn:4p5Compact}, required relatively few exact coefficients 
from perturbation theory once the presence of $\Theta_1(e)$ and $\Xi_1(e)$ was understood and the first part of 
their Taylor expansions was used.  Once the transcendental terms are split off, the fitting methods of 
\cite{ForsEvanHopp16,MunnETC19} could be used to determine the remaining rational series to fairly high order in 
$e^2$.  For the rest of the subleading-log sequence, the same technique might be pushed as high as, say, 15PN, for 
both integer and half-integer in $y$ terms.

However, the integer-order subleading-logs consist of a closed-form part, which appears once the $T_k$ and 
$\Lambda_k$ parts are isolated, as seen with $\mathcal{L}_{6L}$ in \eqref{eqn:sixLogExact}.  Determining 
this entire closed-form part becomes difficult around the 9PN $\log^2$ level, as higher orders in $y$ in BHPT 
calculations require many more decimals of numerical accuracy for a successful PSLQ fit.  Additionally, each 
``jump'' by $y^3 \log(y)$ seems to increase the total number of unknowns, $a_i$ and $b_i$, by 4.  Thus, 
$\mathcal{L}_{9L2}$ would necessitate a fit out to $e^{38}$ to yield an overdetermined system of equations for 
the coefficients in the remaining closed-form terms.  This is no small feat, even using the technique described 
in \cite{MunnETC19} (modified eulerlog procedure) of extracting a purely rational series from each individual flux 
component $\mathcal{L}_{9L2}^{lmn'}$.  Hence, even if determining the entire analytic dependence 
of $\mathcal{L}_{9L2}(e)$ through this method is possible, obtaining the entire eccentricity dependence of any 
further integer-order subleading-logs in the sequence would be prohibitively expensive through fitting alone.

However, there does exist an alternate way forward, which allows for an easier calculation of complicated 
high-PN logarithms like $\mathcal{L}_{9L2}(e)$ to high (finite) order in $e^2$.  In a private communication,
Nathan Johnson-McDaniel revealed a means by which his circular-orbit $S_{lm}$ tail factorization 
\cite{JohnMcDa14} (based on earlier work in \cite{DamoNaga07,DamoIyerNaga09}) 
can be extended to an $S_{lmn'}$ tail factorization for eccentric orbits.  This $lmn'$ factorization
can be combined with fitting methods to greatly simplify (relative to fitting alone) the process of computing 
certain logarithmic PN terms to arbitrary order in $e^2$.  Interestingly, the log terms which can be obtained in 
this manner include the first five PN corrections to \textit{any} integer-order leading logarithm and the first 
four PN corrections to \textit{any} half-integer-order leading logarithm.  As a result, subleading logarithms 
can be determined using this approach. 

This procedure begins by picking a desired order $p$ for corrections to the leading logarithms.  For example, since 
the subleading-log terms addressed in this section are 3PN corrections to the prior leading-log term, to consider 
subleading logs we need to take $p=3$.  Then, secondly, we pick a desired order $\alpha$ in the eccentricity 
expansion (i.e., having the expansion stop at $e^{2 \alpha}$).  Next, the exact analytic form must be found of all 
the $lmn'$ modes needed to reach $y^{p}$ (relative order) in the full flux with an eccentricity expansion to 
$e^{2 \alpha}$.  This can be done by either fitting high-precision numerical data or by direct analytic expansion 
of the equations of BHPT \cite{KavaOtteWard15,HoppKavaOtte16}.  (Indeed, we have begun to supplement numerical 
results with output from a newly-written \textsc{Mathematica} code that does the PN expansions symbolically and 
outputs analytic PN expressions.)  Either way this will produce expressions for a total of approximately 
$2 \alpha \lceil {(p^2 + 6 p + 3)/2} \rceil$ modes.  Each individual $lmn'$ mode is then subjected to tail 
factorization using $S_{lmn'}$ and re-expanded, which removes the transcendentals and leaves a rational double 
expansion through $y^{p}$ and $e^{2 \alpha}$.  Note in the example of $p=3$, this leaves an expansion in rationals 
only through 3PN ($y^3$).  In the next step, we expand each $S_{lmn'}$ tail factor to an \textit{arbitrary order} in 
$y$ and $e^2$.  Then the expanded $lmn'$ tail factors are multiplied by the rational series expansions for $lmn'$, 
re-expanded, and summed over all modes.  The result, remarkably, generates \textit{all} members of the $(p)$PN 
correction to the leading-logarithm series to $e^{2 \alpha}$.  Again, in the $p=3$ example, once we have all modes 
necessary to reach 3PN in the (relative) flux in fully analytic form, expansion of the $S_{lmn'}$ to high PN order 
provides everything we need to find all the subleading logs, e.g., 6L, 9L2, 12L3, etc., to high PN order.

In the particular example of subleading-log $\mathcal{L}_{9L2}(e)$, factored $lmn'$ modes have to be analytically 
calculated up to $l=5$, $m=5$ (excluding 50, 52, and 54) in order to reach 3PN order, and 38 $n'$ modes are needed to 
reach $e^{38}$ for each $lm$.  Multiplying each such mode by the analytic expansion of its respective $S_{lmn'}$, 
with the analytic expansion carried to 9PN order, and then summing all modes together will yield (among other 
things) $\mathcal{L}_{9L2}(e)$ to $e^{38}$.  Those results can then be combined with the Newtonian mass quadrupole 
sums $T_3(e)$ and $\Lambda_3(e)$ to produce a compact, $\mathcal{L}_{6L}(e)$-type \eqref{eqn:sixLogExact} 
solution for $\mathcal{L}_{9L2}(e)$.  Finally, $\mathcal{L}_{9L2}(e)$ can be coupled with $\mathcal{L}_{6L2}(e)$,
$\mathcal{L}_{7L2}(e)$, and $\mathcal{L}_{8L2}(e)$ (listed in \cite{MunnETC19}), along with \eqref{eqn:eToet}, to 
obtain $\mathcal{R}_{9L2}(e_t)$.

Despite the added cost of symbolic calculation, Johnson-McDaniel's $lmn'$ factorization provides a significant
computational speedup over fitting alone, particularly when attempting to reach high order in $y$.  Additionally,
setting $p=0$ in the above procedure reveals an alternative means of calculating the leading logarithms themselves 
to arbitrary order in $e^2$.  By setting $p=0$, we only require an analytic expansion of the $lm$ modes needed 
to give the Peters-Mathews flux (i.e., $l=2,m=-2,0,2$) with the range in $n'$ determined by the desired expansion 
in $e^2$.  The $S_{lmn'}$ factors are then expanded for this more restricted number of modes and used in the 
procedure above.  We have used it to verify the results of Sec.~\ref{sec:entireLL} and the given general PN form for 
leading logs out to 21PN ($\mathcal{L}_{21L7}$) in expansions to $e^6$.  Since these terms depend only on the 
Newtonian quadrupole, they convert directly from expansions in $y$ and $e$ to expansions in $x$ and $e_t$ via 
$e\rightarrow e_t$.  Unfortunately, compared to the multipole moment approach, this process becomes increasingly 
expensive at higher powers of $e^2$, where the number of necessary BHPT $lmn'$ modes grows large.  However, 
for the more complicated subleading-log terms like $\mathcal{L}_{6L}, \mathcal{L}_{9L2},$ etc., this factorization 
technique offers an efficient means to generate expansions at high PN order to comparable finite orders in $e^2$.  
Costs will likely be reduced further upon full implementation of direct analytic PN expansion of the BHPT equations.  
Combining that analytic approach with $S_{lmn'}$ factorizations would be additionally fruitful.  

\subsection{More general relations among coefficients in subleading-logarithmic terms}
\label{sec:MoreGeneral}

The preceding subsections described how explicit calculations from perturbation theory can be coupled with 
Newtonian mass quadrupole summations to extract subleading-logarithms, like $\mathcal{R}_{6L}$.  Now, we seek 
to identify some of the broader structure within this sequence of flux terms.  This task will again involve 
complementary discoveries from both perturbation theory and PN theory, meaning most deductions will necessarily 
remain relevant only to lowest order in the mass ratio.  Remarkably, the results will, though, allow for the 
partial delineation of instantaneous and hereditary terms in the flux.  

The process requires analysis of four separate sources of transcendental structure within the flux: 
\begin{enumerate}
\item Fourier tail integrals of the form \cite{ArunETC08a}
\be 
\label{eqn:tailint} 
\int_0^\infty e^{i n \Omega_r \tau} \log^q\left(\frac{\tau}{2 r_0}\right) d\tau, 
\ee
where $q > 0$ is an integer which generally increases with PN order (see, for instance, Eq. (4.8) of 
\cite{MarcBlanFaye16}), $n$ is the same Fourier harmonic number
appearing in $g(n,e_t)$, and $r_0$ is an arbitrary scale parameter that cancels in the full flux.

\item The perturbation theory eulerlog function for $lmn'$ modes (see 
\cite{MunnETC19,DamoIyerNaga09,JohnMcDa14}):
\be 
\label{eqn:eulog} 
\text{eulerlog}_{m,n'}(y) = \gamma_E + \log| 2 m +2 n' | + \frac{1}{2} \log(y) . 
\ee

\item Instantaneous integrals of the form 
\be 
\label{eqn:instLogInt}
\int_0^{2 \pi} \frac{\log^k[(1 - e_t \cos u)/x]}{(1 - e_t \cos u)^{j}} du 
\ee
for integers $(k, j)$, which emerge with various values of $j$ during the orbital average of $\log^{k}(r)$ terms 
in the flux.  We reuse the integer $k$ here to match the index on $T_k$, as we expect the relevant integrals 
(for integer leading/subleading logs) to appear at $(3k)$PN order.  See \cite{ArunETC08b} for a description 
and evaluation of these integrals.

\item The elimination of all divergences as $e_t \rightarrow 1$ (in particular, logarithmic divergences) by using an
expansion in the compactness parameter $1/p$ ($p$ the semi-latus rectum) instead of in $x$ or $y$.

\end{enumerate}

\subsubsection{Comparison of eulerlog functions}

Starting with the first item in the list, we consider the given class of hereditary integrals.  A common 
regularization procedure entails computation of the following integrals:
\be 
\label{eqn:regtailint} 
\int_0^\infty e^{-|n| \alpha \tau} \log^q\left(\frac{\tau}{2 r_0}\right) d\tau , 
\ee
for constant $\alpha$, which is treated as real and positive, but is ultimately replaced by 
$(\text{sign}(-n) i \Omega_r)$ \cite{ArunETC08a, LoutYune17}.  One key facet of these integrals is their 
evaluation yields the transcendentals $\gamma_E$ and $\log(2 |n| \alpha r_0)$ only in the combination 
$(\gamma_E + \log(2 |n| \alpha r_0))^t$ for one or more $t \in \{1,2, \cdots, q\}$.  In fact, we show in 
the Appendix that \eqref{eqn:regtailint} can be calculated by taking the simpler integral 
\be 
\label{eqn:regtailintsimp}  
\frac{1}{|n| \alpha} \int_0^\infty e^{-\tau} \log^q(\tau) d\tau ,
\ee
and transforming the result by $\gamma_E \rightarrow \gamma_E + \log(2 |n| \alpha r_0).$

Once the substitution for $\alpha$ is made and the imaginary portion separated, the transformation becomes 
$\gamma_E \rightarrow \gamma_E +  \pi i/2 \, \text{sign}(-n) + \log(2 |n| \Omega_r r_0)$.  
When products are taken and a sum is made over positive and negative $n$, the relationship between $\pi$ and
the rest of the expression is slightly obscured by the $\text{sign}(-n)$ function; however, the particular linkage 
among the transcendental factors $(\gamma_E + \log(2 |n| \Omega_r r_0))$ must hold everywhere.

This simple connection constitutes a purely hereditary type of eulerlog function.  Taking the Newtonian limit,
assuming some necessary cancellations (see a related discussion in \cite{BiniDamo14a}), and omitting the unphysical 
regularization constant, we obtain a contribution of the form
\be 
B_k \left( \frac{2}{3} \right)^{k-1} \left(\gamma_E + 2 \log(2) + \log \left |\frac{n}{2} \right|
 + \frac{3}{2}\log(x)\right)^k ,
\ee
at (3k)PN order for some constant $B_k$.  
When $k\ge1$, this can be expanded to isolate the two highest powers of $\log(x)$ as
\be 
\label{eqn:tailratio} 
B_k \log(x)^{k-1}\left[ k \left(\gamma_E + 2 \log(2) 
+ \log \left | \frac{n}{2} \right | \right) + \frac{3}{2} \log(x) \right] , 
\ee
thus providing the expected ratio between the highest power of $\log(x)$ and the combination of transcendentals 
that serves as the coefficient for the next highest power of $\log(x)$.

An eccentricity dependence is attached to these tail integrals in the form of time derivatives of
the mass quadrupole (see, for instance, \cite{DamoNaga08, MarcBlanFaye16}).  
One can use a dimensional argument to show that this yields a factor of $(n/2)^{2k} g(n,e_t)$ for integral 
orders (L. Blanchet, private communication).  After adjusting the initial constant to absorb any additional 
rationals, we can sum over $n$ to find that $\log(x)^{k-1}$ must be attached to
\be 
\label{eqn:tailEulCont} 
C_k \left[ \left(k \, \gamma_E + 2k \log(2) + \frac{3}{2} \log(x) \right) T_k +  k \Lambda_k \right] . 
\ee

However, one must again take care to note that \eqref{eqn:tailratio} and \eqref{eqn:tailEulCont} 
only refer to pieces specifically in the hereditary flux.  On the other hand, the $\text{eulerlog}_{m,n'}$ 
function in \eqref{eqn:eulog}, which is derived through BHPT, characterizes the $lmn'$ modes of the 
entire flux.  It is a direct eccentric-orbit extension of the circular-orbit function $\text{eulerlog}_m(x)$ 
presented in \cite{DamoIyerNaga09}.  Then, using a similar argument, we can obtain the following ratio of 
coefficients for $lmn'$ modes in the total flux:
\begin{gather}
\label{eqn:BHPTtailratio}
k \Big(\gamma_E + \log(2) + \log | m + n' | \Big) + \frac{1}{2} \log(x) .
\end{gather}

The $\log| m + n' |$ term will partially contribute to both $(\log(2) \, T_k)$ and $\Lambda_k$ upon summation 
over $lmn'$, obscuring their final coefficients in the flux.  However, $\gamma_E$ and $\log(x)$ must remain fixed 
in the ratio $k$ to $1/2$.  With the leading logarithm series already calculated, the full contribution to the 
leading-log plus subleading-log terms is then found to be 
\be 
\left(-\frac{856}{105}\right)^k \frac{1}{k!} \left(2 k \, \gamma_E + \log(x) \right) T_k(e_t) . 
\ee
Note that if $k=1$, this provides exactly the $\gamma_E$ and $\log(x)$ contributions to the net 3PN flux in 
\eqref{eqn:3PNE}.  Additionally, it is well known that $\gamma_E$ and $\Lambda_k$ are only present in the 
tail---neither makes an appearance in the instantaneous flux.  Therefore, $\Lambda_k$ can be included to get
the full coefficient
\be 
\left(-\frac{856}{105}\right)^k \frac{1}{k!} \left[ \left(2 k \, \gamma_E + \log(x) \right) T_k(e_t)
+ 2 k \Lambda_k(e_t) \right]
\ee

Interestingly, coupling this (full-flux) expression with the tail result \eqref{eqn:tailEulCont} leads to 
another conclusion: the instantaneous portion of the leading logarithm must equal $-(2/3)$ its hereditary 
counterpart, or
\be 
\label{eqn:instLL} 
\mathcal{R}_{(3k)L(k)}^{\rm inst} = -(2/3) \mathcal{R}_{(3k)L(k)}^{\rm tail} =-2 \mathcal{R}_{(3k)L(k)} .  
\ee
 
\subsubsection{Instantaneous connection and logarithmic divergence}

We can move a step further via the last two items on the list.  Expanding out \eqref{eqn:instLogInt} to retain 
the highest two powers of $\log(x)$ leaves
\be 
(-1)^k \log(x)^{k-1} \int_0^{2 \pi} \frac{\log(x) - k\log(1 - e_t \cos u)}{(1 - e_t \cos u)^{j}} du . 
\ee
Multiple integrals like this appear at any particular PN order, differing in values of $j$.  Evaluation and 
summation of all relevant integrals yields (among other terms) a logarithmic portion of the form 
\be
f_k(e_t)\left[ \log(x) - k \log\left(\frac{2(1-e_t^2)}{1+\sqrt{1-e_t^2}}\right) \right] ,
\ee
for some eccentricity function $f_k(e_t)$.  However, \eqref{eqn:instLL} indicates that this instantaneous 
$\log(x)$ must be attached to $-2 \mathcal{R}_{(3k)L(k)}(e_t) $.  Therefore, we must have 
$f_k(e_t) = -2 \mathcal{R}_{(3k)L(k)}(e_t) $.

Finally, we can compile all this information together to determine the following significant portion of 
the subleading-log (3PN log) series:
\begin{widetext}
\begin{gather}
\label{eqn:3PNlogtail}
\mathcal{R}_{(3k)L(k-1)}^{\rm partial} =
\left(-\frac{856}{105}\right)^k \frac{1}{k!}  \left[ \left(2 k \, \gamma_\text{E}+6 k \log(2)
+2 k \log\left(\frac{1-e_t^2}{1+\sqrt{1-e_t^2}} \right) + \log(x)\right)T_{k}(e_t)
+2 k\Lambda_{k}(e_t) \right] ,
\end{gather}
\end{widetext}
for all $k \ge 1$.  A similar expression (with $4\pi$ out front and $T_k \rightarrow \Theta_k$, 
$\Lambda_k \rightarrow \Xi_k$) follows for half-integer terms.  As we can see, the case $k=2$ matches the last line 
of $\mathcal{R}_{6L}$ in \eqref{eqn:R6L}, and we have verified the corresponding portion of $\mathcal{R}_{9L2}$ as 
well.  Moreover, setting $e_t = 0$ for arbitrary $k$ reproduces the known circular-orbit eulerlog ratio, found 
using the BHPT 220 mode.

There is another means by which to confirm the specific relationship among the coefficients of $\log(x)$,
$\log(1-e_t^2),$ and $\Lambda_k(e_t)$ in the above.  As mentioned in the last item on the list, all divergences 
in eccentricity should vanish in a PN expansion that is made over $1/p$ instead of $x$ or $y$.  This includes 
logarithmic divergences like $\log(1-e_t^2)$, which appear in the three listed terms.  Indeed, because $x$ can be 
expanded in $1/p$ as 
\be 
x = \frac{1-e_t^2}{p} + \mathcal{O}(1/p^2) , 
\ee
each power of $\log(x)$ will necessarily contribute a logarithmic divergence as $e_t \rightarrow 1$.  When this 
fact is applied with the divergence of $\Lambda_k(e_t)$ (see Sec.~\ref{sec:generalChi}) we see that the exact ratio 
of coefficients in \eqref{eqn:3PNlogtail} will eliminate all the logarithmic divergences at 
$\log(p)^{k-1}/p^{3k}$ order.  Thus, this alternative fit provides an additional check on our results.

\section{Conclusions and outlook}

This paper has illustrated a relatively novel way to use known BHPT and PN techniques to make progress in 
understanding the PN expansions of the energy and angular momentum gravitational wave fluxes for eccentric-orbit 
EMRIs.  By pairing finite-order eccentricity expansions from BHPT (found either by combining numerical fitting with 
PSLQ or by analytically PN expanding the equations of BHPT directly) with astute predictions for the multipole 
content of select flux terms, we can ascertain exact or greatly simplified forms for the eccentricity dependence of
those terms to high PN orders at lowest order in the mass ratio---results which would otherwise have required 
years of progress in the full PN theory.  In this paper we have shown that several sequences of PN fluxes 
(leading logarithms and subleading logarithms) can be understood in this way merely by seeing the role of the
Newtonian mass quadrupole moment power spectra, $g(n,e_t)$ and $\tilde{g}(n,e_t)$.

More specifically, we showed in Sec.~\ref{sec:entireLL} that the entire sequence of integer in $x$ PN-order 
leading-log terms are closed-form expressions in $e_t$ and the entire sequence of 
half-integer in $x$ leading-log terms are infinite series in $e_t^2$ with easily determined rational coefficients.  
For the energy flux, the Newtonian mass quadrupole moment enters into these sequences of terms through the Fourier 
sums $T_k(e_t)$ and $\Theta_k(e_t)$, which are sums over filtered weightings of the quadrupole spectrum $g(n,e_t)$.  
Equivalent sums exist for leading-log angular momentum fluxes.

Yet the Newtonian mass quadrupole moment plays an even wider role than just explaining the leading-log sequences.  
As Sec.~\ref{sec:additional6L} showed, adequate BHPT results can in principle be combined with an ansatz for how 
the Newtonian quadrupole moment enters the subleading-log flux sequences to determine completely their eccentricity 
dependence also.  With the subleading-log sequences, two new sets of Fourier sums, $\Lambda_k(e_t)$ and 
$\Xi_k(e_t)$, are defined from the quadrupole spectrum $g(n,e_t)$ (with mirror images for angular momentum).  We 
then demonstrated the process explicitly with the (integer-order) $\mathcal{R}_{6L}(e_t)$ flux term.  At 
half-integer in $x$, adequate BHPT data and essentially the same procedure also allowed a key decomposition of the 
subleading-log term $\mathcal{L}_{9/2}(e)$, revealing in that case an infinite series in $e^2$ with rational 
coefficients that can be determined to high order in $e^2$.  We suspect that this procedure can be applied 
successfully to higher PN order subleading-log terms, giving complete $\mathcal{R}_3$-type analytic 
representations for $\mathcal{L}_{9L2}(e)$, $\mathcal{L}_{12L3}(e)$, etc., and their $\mathcal{R}_i(e_t), 
\mathcal{J}_i(e), \mathcal{Z}_i(e_t)$ counterparts.  We also suspect that $\mathcal{L}_{9/2}$-type segregations of 
transcendental terms and rational-coefficient infinite series will occur at higher PN orders for all half-integer in 
$x$ subleading-logs, like $\mathcal{L}_{15/2}(e)$, $\mathcal{L}_{21/2}(e)$, etc., and that these might be found 
given enough BHPT data.

The methods and results developed here are another example in a body of literature using BHPT to inform PN theory 
and vice versa.  Our focus on leading and subleading logarithms, though differing in scope, is strongly reminiscent 
of \cite{JohnMcDaShahWhit15} and \cite{MunnETC19}, who used the appearance of the eulerlog function to develop an 
understanding of lower powers of logarithms from higher ones.  It is also not unlike the calculation of the redshift 
invariant achieved by \cite{BlanETC10}, who combined logarithmic derivations with self-force data to extract 
non-logarithmic terms numerically.

With leading-log and subleading-log fluxes (at lowest order in the mass ratio) so well understood analytically, 
by exploiting the role of the Newtonian mass quadrupole moment spectra and making judicious use of BHPT results, 
what more might be done to find flux terms at high PN order without the full PN formalism?  It turns 
out that similar headway can be made for terms that are a 1PN correction to elements of the leading-log and 
subleading-log sequences (to be reported elsewhere \cite{MunnEvan19b}).  That analysis requires the Fourier 
amplitudes of the next Newtonian multipole moments (current quadrupole and mass octupole) and the 1PN correction 
to the mass quadrupole moment.  Together with the approach of this paper, a pattern emerges for chipping away at 
an analytic understanding of the PN expansion in the fluxes.  Rather than proceed one power in $x$ (or $y$) at a 
time, as would be typical in advances in the full PN formalism, we take each order in multipole moments as a group, 
using them to calculate all the most significant PN contributions from that group.  This leads to making progress 
through the PN expansion in a ``diagonal'' sense.  We first come to understand the eccentricity dependence of the 
entire leading-log (diagonal) sequences, $x^{3k} \log^k(x)$ and $x^{3k+3/2} \log^k(x)$.  Next, we gain an 
understanding of the subleading-log diagonals, with PN dependence $x^{3k} \log^{k-1}(x)$ and 
$x^{3k+3/2} \log^{k-1}(x)$.  Then, as we will show elsewhere \cite{MunnEvan19b}, we can tackle the 1PN corrections 
to the leading-logs, which are the diagonals in the PN expansion with $x^{3k+1} \log^k(x)$ and 
$x^{3k+5/2} \log^k(x)$, and 1PN corrections to the subleading-logs, with $x^{3k+1} \log^{k-1}(x)$ and 
$x^{3k+5/2} \log^{k-1}(x)$.

Stated in different notation, in the subsequent paper on 1PN corrections to leading and subleading logarithms we 
will show additional closed-form expressions for the integer-PN-order 1PN logarithms $\mathcal{R}_{(3k+1)L(k)}(e_t)$ 
and $\mathcal{Z}_{(3k+1)L(k)}(e_t)$ (for $k \ge 0$) (e.g., $\mathcal{R}_{4L}$, $\mathcal{R}_{7L2}$, 
$\mathcal{R}_{10L3}$, etc.) and find infinite power series for half-integer-PN-order 1PN logarithms 
$\mathcal{R}_{(3k+5/2)L(k)}$ and $\mathcal{Z}_{(3k+5/2)L(k)}$ (e.g., $\mathcal{R}_{11/2L}$, $\mathcal{R}_{17/2L2}$, 
etc.), at lowest order in the mass ratio.  Interestingly, there is some prospect that we might ascertain the 
corresponding contributions at next order in $\nu$ as well, though without (at present) second-order BHPT results 
to help in confirmation.  Some of these results have already been obtained simply by PSLQ analysis of high-precision 
BHPT numerical results.  For example, a closed-form expression for $\mathcal{L}_{4L}$ is found in
\cite{ForsEvanHopp16} and other closed-form expressions for $\mathcal{J}_{4L}$, $\mathcal{L}_{7L2}$, and 
$\mathcal{J}_{7L2}$ are found in \cite{MunnETC19}.  Completely new results have been found in making 
1PN corrections to the subleading-logs, with (analytically understood) infinite series obtained for 
$\mathcal{L}_{4}$ and $\mathcal{J}_{4}$ \cite{MunnEvan19b}.  The remaining integer-order 1PN corrections to 
the subleading-logarithms (e.g., $\mathcal{L}_{7L}$, $\mathcal{L}_{10L2}$, etc.) can be similarly obtained by 
combining the other 1PN logarithms with the $S_{lmn'}$ factorization.  The irrational portions of half-integer-order 
terms like $\mathcal{L}_{11/2}$, $\mathcal{L}_{17/2L}$, $\mathcal{L}_{23/2L2}$, etc., will likely follow as well.  

To provide a more concrete view of how all these pieces tie together, Table \ref{tab:PNterms} shows the present 
state of knowledge of the eccentricity dependence of energy flux terms $\mathcal{L}_i(e)$ for PN orders through 
7.5PN order and (somewhat) beyond, at lowest order in the mass ratio.  Analogous depth of understanding exists 
for the angular momentum fluxes, $\mathcal{J}_i(e)$.  Going beyond these orders, converting to $\mathcal{R}_i(e_t)$ 
and $\mathcal{Z}_i(e_t)$, and moving to higher orders in $\nu$ are all subjects for potential future work.

\begin{widetext}
\begin{center}
\begin{table}[h!]
\label{tab:PNterms}
\parbox{17.9cm}{
\caption{State of knowledge of eccentricity dependence of PN flux terms.  The second column is the power series 
expansion order in $e$ to which the respective flux term is known at present.  The terms $\mathcal{L}_3$ and 
$\mathcal{L}_{3L}$ were previously known \cite{ArunETC08a}.  The closed-form result for $\mathcal{L}_{4L}$ was 
also previously known \cite{ForsEvanHopp16}.  All other results come from this paper and its companion 
\cite{MunnETC19}.  Flux terms labeled as ``all orders" are infinite series in $e^2$ with analytically 
calculable coefficients.  Other terms are ``only'' known in analytic form up to order $e^{30}$ (or in a few cases 
less).  The third column gives the number of PN corrections to the leading-logs which must be calculated 
to derive the term fully.  The fourth column indicates the number of leading log (and $\Lambda(e_t)/\Xi(e_t)$) 
corrections which must be calculated to extract the term to all orders in $e$ in the manner of 
Sec.~\ref{sec:additional6L}.  A superset of these terms allow for the separation of transcendental contributions 
in the same way, as shown in column five.  Above 5PN it is more difficult to apply these methods (labeled by 
asterisk).  The last two rows represent all further leading logarithms.\\}}
\begin{tabular}{ || c | c | c | c | c ||}
\hline\hline
Term & Known order in $e$ & PN Order beyond LL & Order for fitting extraction & Order to find transcendental part \\
\hline
$\mathcal{L}_{3}$ &\cellcolor{GreenYellow} All Orders & 3PN & 0PN & 0PN \\ 
\hline
$\mathcal{L}_{3L}$ &\cellcolor{lime} Closed Form & --- & --- & --- \\
\hline
$\mathcal{L}_{7/2}$ &\cellcolor{Yellow} Fitted to $e^{30}$ & 2PN & --- & ---  \\
\hline
$\mathcal{L}_{4}$ &\cellcolor{Yellow} Fitted to $e^{30}$ & 4PN & 1PN & 1PN \\
\hline
$\mathcal{L}_{4L}$ &\cellcolor{lime} Closed Form & 1PN & --- & --- \\
\hline
$\mathcal{L}_{9/2}$ &\cellcolor{Yellow} Fitted to $e^{30}$ & 3PN & --- & 0PN \\
\hline
$\mathcal{L}_{9/2L}$ &\cellcolor{GreenYellow} All Orders & --- & --- & --- \\
\hline
$\mathcal{L}_{5}$ &\cellcolor{Yellow} Fitted to $e^{30}$ & 5PN & 2PN & 2PN \\
\hline
$\mathcal{L}_{5L}$ &\cellcolor{lime} Closed Form & 2PN & --- & --- \\
\hline
$\mathcal{L}_{11/2}$ &\cellcolor{Yellow} Fitted to $e^{30}$ & 4PN & --- & 1PN \\
\hline
$\mathcal{L}_{11/2L}$ &\cellcolor{Yellow} Fitted to $e^{30}$ & 1PN & --- & --- \\
\hline
$\mathcal{L}_{6}$ &\cellcolor{YellowOrange} Fitted to $e^{20}$ & 6PN & 3PN* & 3PN* \\
\hline
$\mathcal{L}_{6L}$ &\cellcolor{GreenYellow} All Orders & 3PN & 0PN & 0PN \\
\hline
$\mathcal{L}_{6L2}$ &\cellcolor{lime} Closed Form & --- & --- & --- \\
\hline
$\mathcal{L}_{13/2}$ &\cellcolor{Yellow} Fitted to $e^{30}$ & 5PN & --- & 2PN \\
\hline
$\mathcal{L}_{13/2L}$ &\cellcolor{Yellow} Fitted to $e^{30}$ & 2PN & --- & --- \\
\hline
$\mathcal{L}_{7}$ &\cellcolor{Orange} Fitted to $e^{12}$ & 7PN & 4PN* & 4PN* \\
\hline
$\mathcal{L}_{7L}$ &\cellcolor{YellowOrange} Fitted to $e^{26}$ & 4PN & 1PN & 1PN \\
\hline
$\mathcal{L}_{7L2}$ &\cellcolor{lime} Closed Form & 1PN & --- & --- \\
\hline
$\mathcal{L}_{15/2}$ &\cellcolor{Orange} Fitted to $e^{12}$ & 6PN & --- & 3PN* \\
\hline
$\mathcal{L}_{15/2L}$ &\cellcolor{YellowOrange} Fitted to $e^{26}$ & 3PN & --- & 0PN \\
\hline
$\mathcal{L}_{15/2L2}$ &\cellcolor{GreenYellow} All Orders & --- & --- & --- \\
\hline
$\mathcal{L}_{(3k)L(k)}$ &\cellcolor{lime} Closed Form & --- & --- & --- \\
\hline
$\mathcal{L}_{(3k)L(k+3/2)}$ &\cellcolor{GreenYellow} All Orders & --- & --- & --- \\
\hline \hline
\end{tabular}
\end{table}
\end{center}

\acknowledgments

We thank Nathan Johnson-McDaniel and Luc Blanchet for helpful 
discussions.  This work was supported by the North Carolina Space Grant.  
This work was also supported in part by NSF grants PHY-1506182 and 
PHY-1806447.  C.R.E.~acknowledges support from the Bahnson Fund at the 
University of North Carolina-Chapel Hill.  

\begin{appendix}

\section{The tail eulerlog function}

We prove that the integral \eqref{eqn:regtailint} can be found using
the simpler integral \eqref{eqn:regtailintsimp} under the transformation 
$\gamma_E \rightarrow \gamma_E + \log(2 |n| \alpha r_0)$.  To proceed, we first write general forms for the two 
integrals.  Recall that the Gamma function $\Gamma(x)$ is given by
\be \Gamma(x) = \int_0^{\infty} t^{x-1} e^{-t} dt. \ee
Then, the two tail integrals can be written as \cite{GradETC07}
\begin{gather}
\int_0^{\infty}  e^{-\tau} \log^{q} (\tau) d\tau 
= \frac{d^{q} \Gamma(x + 1)}{dx^{q}}  \bigg |_{x = 0},  \notag \\
\int_0^{\infty}  e^{-|n| \alpha \tau} \log^{q} \left(\frac{\tau}{2 r_0}\right) d\tau 
 = \frac{1}{|n| \alpha} \frac{d^{q}}{dx^{q}} 
 \left( \frac{\Gamma(x + 1)}{(2 |n| \alpha r_0)^{x}} \right) \bigg |_{x = 0},
 \label{eqn:integralsGamma}
\end{gather}
where the second relation is obtained using the variable substitution $u = |n| \alpha \tau$.

Both $\Gamma(x+1)$ and $\Gamma(x+1)/(2 |n| \alpha r_0)^{x}$ permit convergent Taylor series about 
$x = 0$ for $|x| < 1$.  These are most easily computed using the following representations,
valid for $|x| < 1$ \cite{NHMF}:
\begin{gather}
\Gamma(x+1) = \exp\left(-\gamma_E x + \sum_{k=2}^{\infty} \frac{\zeta(k)}{k} (-x)^k \right) \notag \\
\frac{\Gamma(x+1)}{(2 |n| \alpha r_0)^{x}} = 
(2 |n| \alpha r_0)^{-x} \exp\left(-\gamma_E x + \sum_{k=2}^{\infty} \frac{\zeta(k)}{k} (-x)^k \right) 
\label{eqn:gammaSer}
\end{gather}
Then, either integral containing $\log^{q}$ can be calculated by expanding
the necessary term about $x = 0$ and picking out the coefficient
of $x^{q}/q!$, possibly with a factor of $1/|n|\alpha$.  
But the second expression can be rewritten as
\begin{gather}
\frac{\Gamma(x+1)}{(2 |n| \alpha r_0)^{x}} = 
\exp\left(-(\gamma_E +\log(2 |n| \alpha r_0) )x + \sum_{k=2}^{\infty} \frac{\zeta(k)}{k} (-x)^k \right). 
\end{gather}
Thus, this latter series can be evaluated by making the 
substitution $\gamma_E \rightarrow \gamma_E + \log(2 |n| \alpha r_0)$ in the first.  
This completes the proof.

The above results imply a way in which $g(n,e_t)$ appears in $\mathcal{R}_6$.  Given the form of the exponentials 
in \eqref{eqn:gammaSer}, it seems likely that the hereditary flux will source the appearance of certain 
transcendentals like $\zeta(3)$ at higher orders in the PN expansion.  Indeed, we can see in the BHPT fitting 
results from \cite{MunnETC19, BHPTK18} that the 6PN term $\mathcal{L}_6$ contains three such pieces:
\be
\mathcal{L}_6^{\rm partial} = - \left( \frac{27392}{105} \zeta(3)
+ \frac{256}{45} \pi^4  
+ \frac{27392}{315} \gamma_E \, \pi^2  \right) \, T_2(e) .
\ee
Of course, because the eccentricity dependence is solely determined by the Newtonian sum $T_2$,
$\mathcal{R}_6(e_t)$ will have the same three contributions with $T_2(e) \rightarrow T_2(e_t)$.
\end{appendix}
\end{widetext}

\bibliography{leadinglogs}

\end{document}